\documentclass[prd,twocolumn,superscriptaddress,amsmath,amssymb,floatfix,preprintnumbers,10pt,nofootinbib,floatfix]{revtex4-1}
\usepackage{booktabs,color,graphicx,verbatim}
\usepackage[colorlinks=true,citecolor=blue,filecolor=blue,linkcolor=blue,urlcolor=blue,pdftex]{hyperref}
\newcommand{\Sone}{\mathrm{S1}}
\newcommand{\Stwo}{\mathrm{S2}}
\newcommand{\mN}{m_{\mathrm{T}}}
\newcommand{\muN}{\mu_{\mathrm{T}}}
\newcommand{\mDM}{m_{\mathrm{DM}}}

\newcommand{\amsterdam}{\affiliation{GRAPPA Centre of Excellence, Institute for Theoretical Physics Amsterdam and Delta Institute for Theoretical Physics, University of Amsterdam, Science Park 904, 1098 XH Amsterdam, The Netherlands}}
\newcommand{\london}{\affiliation{Department of Physics, King's College London, Strand, London, WC2R 2LS, United Kingdom}}

\begin{document}

\preprint{KCL-PH-TH/2017-16}
\title{New constraints and discovery potential of sub-GeV dark matter with xenon detectors}

\author{Christopher McCabe}
\email{christopher.mccabe@kcl.ac.uk}
\amsterdam
\london

\begin{abstract}
Existing xenon dark matter (DM) direct detection experiments can probe the DM-nucleon interaction of DM with a sub-GeV mass through a search for photon emission from the recoiling xenon atom.  We show that LUX's constraints on sub-GeV DM, which utilise the scintillation (S1) and ionisation (S2) signals, are approximately three orders of magnitude more stringent than previous xenon constraints in this mass range, derived from the XENON10 and XENON100 S2-only searches. The new LUX constraints provide the most stringent direct detection constraints for DM particles with a mass below~0.5~GeV. In addition, the photon emission signal in LUX and its successor LZ maintain the discrimination between background and signal events so that an unambiguous discovery of sub-GeV DM is possible. We show that LZ has the potential to reconstruct the DM mass with $\simeq20\%$ accuracy for particles lighter than~0.5~GeV. 
\end{abstract}

\maketitle

\section{Introduction}

Identifying the nature of dark matter (DM) remains one of the most compelling problems in astroparticle physics. Motivated by the weakly interacting massive particle (WIMP) paradigm, DM direct detection experiments have traditionally concentrated on the 5~GeV to 10~TeV mass range. The non-detection of DM in this range has led to significant theoretical efforts focussing on lighter particles. Initially, sub-GeV DM was mainly considered in the context of the 511~keV gamma-ray anomaly observed by INTEGRAL~\cite{Boehm:2003hm, Boehm:2003bt,Boehm:2003ha}, but more recently, it has been recognised that sub-GeV DM is generic in many other scenarios, e.g.~\cite{Pospelov:2007mp,Hall:2009bx, Hochberg:2014dra,DAgnolo:2015ujb,Pappadopulo:2016pkp,DAgnolo:2017dbv}.

This theoretical activity has motivated both new direct detection experiments for sub-GeV DM~\cite{Essig:2015cda,Hochberg:2015fth,Hochberg:2016ntt,Derenzo:2016fse,Hochberg:2016ajh, Hochberg:2016sqx,Bunting:2017net,Guo:2013dt,Schutz:2016tid, Knapen:2016cue,Essig:2016crl}, and new searches with existing experiments~\cite{Pospelov:2008jk,Essig:2011nj,Essig:2012yx,An:2014twa,Lee:2015qva,Bloch:2016sjj}. The major obstacle faced by low mass searches is that the energy deposited in a detector by sub-GeV DM is small. For instance, the maximum recoil energy imparted by DM (of mass $\mDM$) to a nucleus (with mass number~$A$) in elastic scattering is $E^{\rm{max}}_{\rm{R}}\approx 0.1~\mathrm{keV} \cdot  (131/A)\, (\mDM/1~\mathrm{GeV})^2$. The nuclear recoil energy threshold of dual-phase xenon detectors $(A_{\rm{Xe}}\simeq131)$ is approximately $1$~keV, implying that they are limited to $\mDM\gtrsim3$~GeV. The lighter nuclei and lower energy thresholds employed in the CRESST~\cite{Angloher:2015ewa, Angloher:2015eza}, DAMIC~\cite{Aguilar-Arevalo:2016ndq}, EDELWEISS~\cite{Hehn:2016nll} and SuperCDMS~\cite{Agnese:2015nto,Agnese:2016cpb}  detectors allow them to probe lower masses, with current exclusion limits reaching down to $\mDM\simeq 0.5$~GeV. Unfortunately, the push to a lower energy threshold often comes with the loss of good discrimination between background and DM events, limiting their ability to make an unambiguous discovery of sub-GeV~DM. 
 
Reference~\cite{Kouvaris:2016afs} demonstrated that existing xenon detectors can probe sub-GeV DM through a new signal channel: a search for the irreducible photon emission from a polarised xenon atom, caused by the displacement of the nucleus and electron charges after the xenon nucleus recoils, and derived constraints from the XENON10 and XENON100 S2-only searches~\cite{Angle:2011th,Aprile:2016wwo}. In this $\mathrm{DM}+\mathrm{Xe}\to\mathrm{DM}+\mathrm{Xe}+\gamma$ inelastic scattering process with a photon in the final state, the maximum photon energy is $\omega^{\rm{max}}\approx 3~\mathrm{keV} \cdot (\mDM/1~\mathrm{GeV})$. In this paper, we show that the LUX dual-phase xenon detector can also probe sub-GeV DM with the more powerful $\mathrm{S1}+\mathrm{S2}$ search, which is sensitive to photon energies $\omega\gtrsim0.3$~keV.\footnote{The~S1 and~S2 signals are defined more carefully in Sec.~\ref{sec:LUXLZ}.} We calculate the parameter space excluded with data from LUX's two WIMP searches (WS2013~\cite{Akerib:2015rjg} and WS2014-16~\cite{Akerib:2016vxi}) and show that the LUX constraints are up to three orders of magnitude more sensitive than the XENON10 and XENON100 S2-only searches considered in Ref.~\cite{Kouvaris:2016afs}. This is because the photon's energy is large enough to produce detectable scintillation and ionisation charge signals, with the result that events from the fiducial volume, where the background is lower~\cite{Undagoitia:2015gya}, can be selected.  Moreover, the good discrimination between background and signal events based on the scintillation and ionisation signals is retained. This further reduces the background rate and importantly, allows for an unambiguous detection of DM to be made. We demonstrate this explicitly for LZ~\cite{Akerib:2015cja}, where we calculate its sensitivity and show that an experiment under construction has the potential to accurately reconstruct the parameters of sub-GeV~DM.

\section{Photon emission scattering rate}

The differential rate for a DM particle to undergo two-to-three scattering with a nucleus of mass $\mN$ is
\begin{equation}
\frac{d R}{ d \omega}= \frac{\rho_{\mathrm{DM}}}{\mN \mDM}  \int_{v_{\rm{min}}}\!\!\!\! d^3 v v f(\mathbf{v}+\mathbf{v}_{\rm{E}})  \int_{E^-_{\rm{R}}}^{E^+_{\rm{R}}} \!\!\!    d E_{\rm{R}} \frac{d^2 \sigma}{d \omega d E_{\rm{R}}},
\end{equation}
where $\rho_{\rm{DM}}=0.3~\mathrm{GeV}/{\mathrm{cm}}^3$ is the local DM density and $f(\mathbf{v})$ is the DM velocity distribution in the Galactic frame, which we assume is a Maxwell-Boltzmann (MB) distribution with a cut-off at $v_{\rm{esc}}=544~\mathrm{km}/\mathrm{s}$ and most probable speed of $v_0=220~\mathrm{km}/\mathrm{s}$. We boost from the Galactic to the Earth reference frame with~$\mathbf{v}_{\rm{E}}$~\cite{McCabe:2013kea,Lee:2013xxa}. Small deviations from a MB distribution are likely, as seen in numerical simulations, e.g.~\cite{Bozorgnia:2016ogo,Kelso:2016qqj,Sloane:2016kyi}, and predicted by Earth-scattering effects, e.g.~\cite{Collar:1992qc, Collar:1993ss,Kavanagh:2016pyr}, but we do not consider them in this work. The limits of integration are found from three-body kinematics, $E_{\rm{R}}^{\pm}= \muN^2 v^2/\mN \cdot \left[1-v_{\rm{min}}^2/(2v^2) \pm \sqrt{1- v_{\rm{min}}^2/v^2} \right]$, where $v_{\mathrm{min}}=\sqrt{2 \omega/\muN}$ is the minimum DM speed required for a photon to have energy~$\omega$, and $\muN$ is the DM-nucleus reduced mass.

The photon emission cross-section (derived in Ref.~\cite{Kouvaris:2016afs})~is
\begin{equation}
\frac{d^2 \sigma}{d \omega d E_{\rm{R}}}=\frac{4 \alpha}{3 \pi}\frac{\left| f(\omega)\right|^2 }{ \omega} \frac{E_{\rm{R}}}{\mN}  \frac{d \sigma}{ d E_{\rm{R}}}\;,
\end{equation}
where $\alpha$ is the fine-structure constant, $f(\omega)=f_1 (\omega)+i f_2 (\omega)$ are atomic form factors~\cite{atomicff}, and $d \sigma/d E_{\rm{R}}$ is the DM-nucleus cross-section for elastic scattering. The price to pay for photon emission is a factor $E_{\rm{R}}/\mN$, resulting in a $\mathcal{O}(0.1~\mathrm{keV}/100~\mathrm{GeV})\simeq\mathcal{O}(10^{-9})$ suppression factor.

We parameterise the DM-nucleus cross-section as
\begin{equation}
\frac{d \sigma}{d E_{\rm{R}}} = \frac{\mN\, \sigma^0_{\rm{SI}}}{2 \mu_n^2 v^2} \,F^{\rm{SI}}_{\rm{T}}(E_{\rm{R}}) \,F_{\rm{med}} (E_{\rm{R}})\;,
\end{equation}
where $\mu_n$ is the DM-nucleon reduced mass and $F^{\rm{SI}}_{\rm{T}}$ is the nuclear form factor. It is an excellent approximation to evaluate $F^{\rm{SI}}_{\rm{T}}$ at $E_{\rm{R}}=0$~keV. We focus on spin-independent (SI) interaction with equal interaction strength with protons and neutrons so $F^{\rm{SI}}_{\rm{T}}=A^2$~\cite{Vietze:2014vsa}. Finally, $F_{\rm{med}}(E_{\rm{R}})$ is a factor that depends on the mass of the particle mediating the interaction. In the heavy mediator limit, $m_{\rm{med}}\gg q$, where $q= \sqrt{2 \mN E_{\mathrm{R}}}\sim3~\mathrm{MeV}\cdot (\mDM/1~\mathrm{GeV})$ is the momentum transfer, $F_{\rm{med}}=1$ and hence, $\sigma^0_{\rm{SI}}$ is the usual DM-nucleon cross-section that is constrained in SI analyses. In the light mediator limit, $m_{\rm{med}}\ll q$, $F_{\rm{med}}=q_{\rm{ref}}^4/q^4$. In this case, $\sigma^0_{\rm{SI}}$ must be defined at a reference value of $q$; we take $q_{\rm{ref}}=1~\mathrm{MeV}$, the typical size of $q$ for $\mDM\lesssim1$~GeV.

\section{LUX and LZ detector simulations \label{sec:LUXLZ}} 

Dual-phase xenon detectors do not directly measure energy. Rather, they measure the `S1' and `S2' signals, proportional to the initial scintillation and ionisation charge respectively, produced by an energy deposition~\cite{Chepel:2012sj}.

The event rate in terms of the observable signals is
\begin{equation}
\frac{d^2R}{d\Sone\,d\Stwo} =\epsilon(\Sone,\Stwo)\int d \omega \, \frac{d R}{ d \omega} \,\mathrm{pdf}(\Sone,\Stwo|\omega)\;,
\end{equation}
where $\epsilon(\Sone,\Stwo)$ represent detection efficiencies and we determine $\mathrm{pdf}(\Sone,\Stwo|\omega)$ with a Monte Carlo simulation of the detector. Our simulations are based on the Noble Element Simulation Technique (NEST)~\cite{Szydagis:2011tk,Szydagis:2013sih,Lenardo:2014cva} and following Ref.~\cite{Szydagis:2013sih}, we assume that the electron and photon yields, $Q_y$ and $L_y$ respectively, from keV-energy depositions from beta particles and gamma rays are the same (collectively, electronic recoils). For an energy $E$, the mean~S1 and~S2 values are related to the yields through $\mathrm{S1}=g_1 L_y E$ and $\mathrm{S2}=g_2 Q_y E$, where $g_1$ and $g_2$ are proportionality (or `gain') factors. Following theoretical arguments, we assume that the $\mathcal{O}(0.1)~\mathrm{keV}$ nuclear recoil associated with the $\mathrm{DM}+\mathrm{Xe}\to\mathrm{DM}+\mathrm{Xe}+\gamma$ inelastic scattering process does not produce an observable signal~\cite{Sorensen:2014sla}. There are proposals to test this assumption with new low-energy calibration techniques~\cite{Verbus:2016sgw}.

First, we describe our simulation for electronic recoils (ERs), where our input is $Q_y$. Above 1.3~keV, we fit~$Q_y$ to the central values of LUX's tritium calibration data~\cite{Akerib:2015wdi}. Below 1.3~keV, we fit to the central values from LUX's calibration with $^{127}\mathrm{Xe}$~\cite{xe127}. There are no calibration data below 0.19~keV so we assume that~$Q_y$ is zero below this energy. We self consistently determine~$L_y$ through the relation $n_q/E=Q_y+L_y$, where $n_q=E/13.7~\mathrm{eV}$ is the total number of quanta from an energy~$E$.
Our~$L_y$ agrees perfectly with LUX's~$L_y$ calibration data above 1.3~keV~\cite{Akerib:2015wdi}. As with~$Q_y$, we assume that~$L_y$ is zero below 0.19~keV. Our yields are also in good agreement with data from the PIXeY xenon detector~\cite{BoultonAPS,Boulton:2017hub}. We include recombination fluctuations, generating the recombination probability~$r$ and the fluctuations as described in Ref.~\cite{Akerib:2016qlr} with parameters $\sigma_p=0.07$ and~$\alpha=0.2$. Additional parameterisations of~$Q_y$ are investigated in Appendices~\ref{sec:signal} and~\ref{sec:effects}.

Second, we summarise our simulation for nuclear recoils (NRs). This is used to check that our simulations correctly reproduce published LUX NR results and also to calculate the $^8\rm{B}$ solar neutrino signal in~LZ. As input, we use a~$Q_y$ and~$L_y$ parameterisation that fits LUX's D-D calibration data~\cite{Akerib:2016mzi} and assume the Lindhard model with $k=0.174$~\cite{Akerib:2016mzi}. We include Penning quenching as in Ref.~\cite{Akerib:2016mzi}, model the recombination probability following the Thomas-Imel model with parameters in Ref.~\cite{Bailey}, and again use~$\sigma_p=0.07$ to model recombination fluctuations. 

Unless stated otherwise, S1 and S2 refer to position corrected values, where S1 is normalised to the centre of the detector and S2 to the top of the liquid. We take into account the variation of $g_1$ with height within the detector (we ignore radial variations) using results in Ref.~\cite{Dobi} for LUX and projections for LZ in Ref.~\cite{Akerib:2015cja}. For LUX~(LZ), we use an electron lifetime of 800~(3000)~ms and an electron drift speed 1.5~mm/$\mu$s for both. 

For LUX~WS2013, we use the parameter values from Refs.~\cite{Akerib:2015rjg,Akerib:2016lao}: $g_1=0.117~\mathrm{phd}/\gamma$, $g_2=12.1~\mathrm{phd}/e^-$, an extraction efficiency of $49\%$; the~S1 detection efficiency from Ref.~\cite{Bailey}; we allow events that satisfy $\mathrm{S1}_{\mathrm{raw}}>1~\mathrm{phd}$, $\mathrm{S2}_{\rm{raw}}>165~\mathrm{phd}$; and compare against events measured with $\mathrm{radius}<18$~cm to set an exclusion limit.

\begin{figure}[!t]
\centering
\includegraphics[width=0.95\columnwidth]{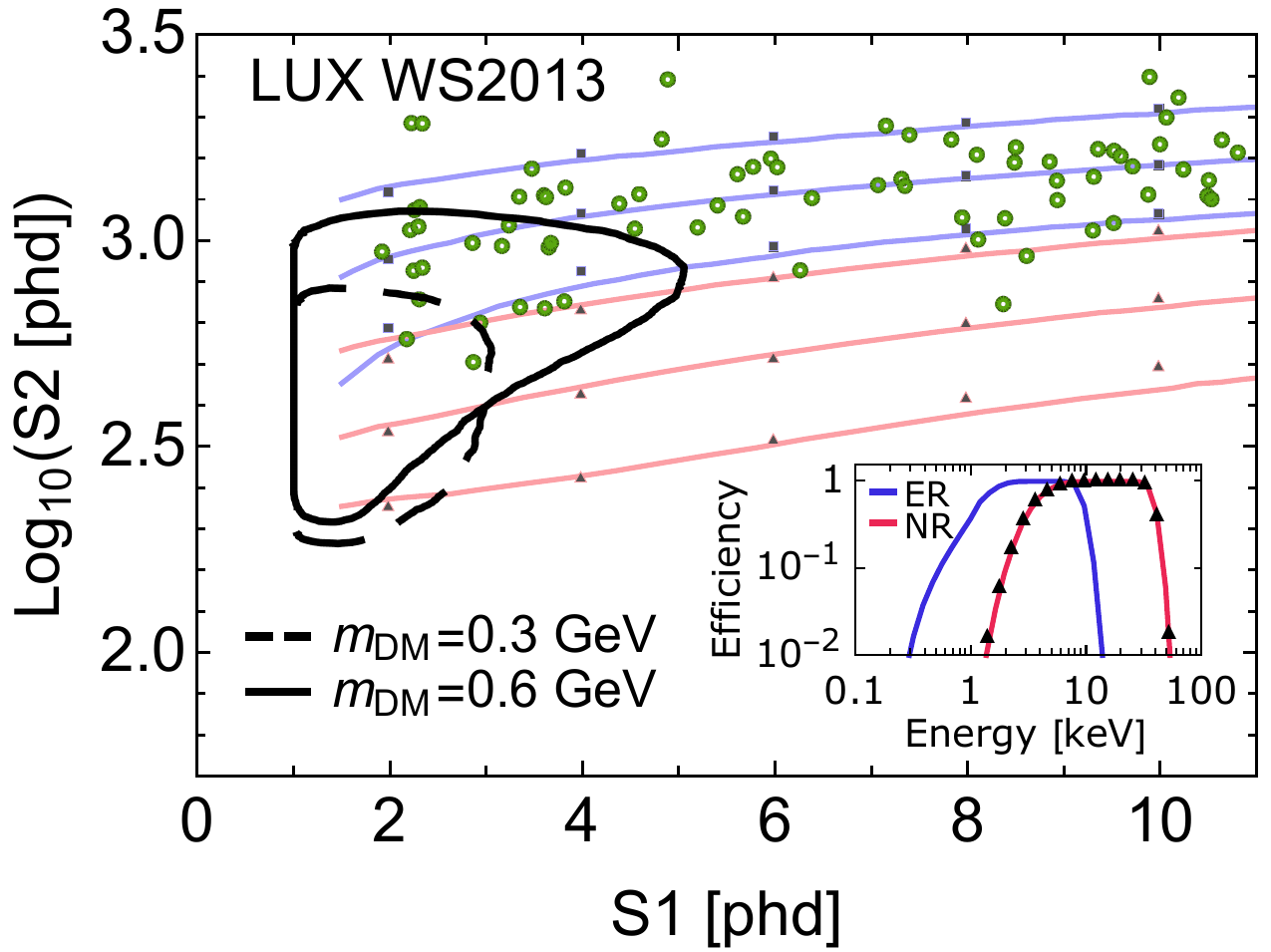}\\
\includegraphics[width=0.95\columnwidth]{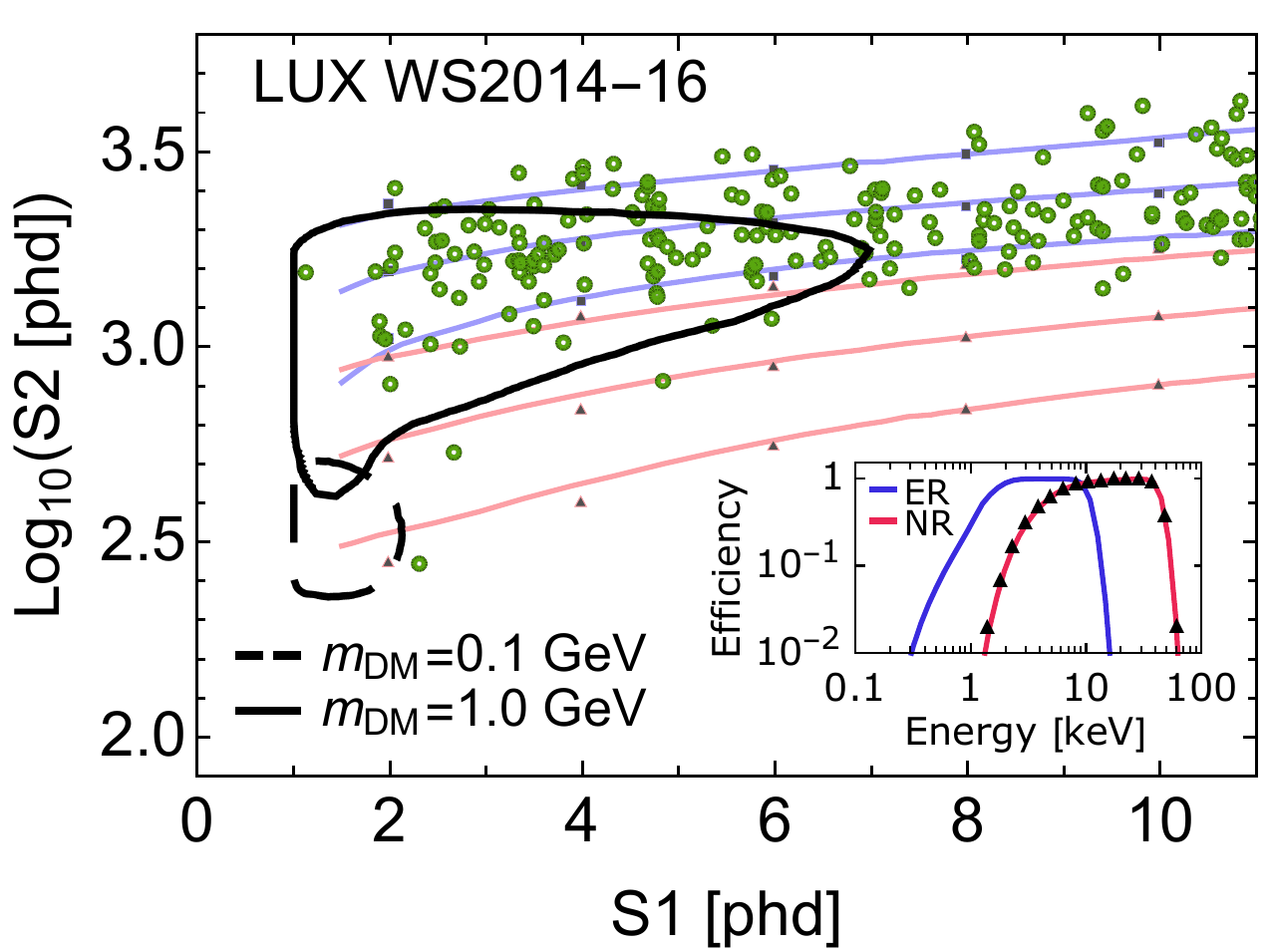}
\caption{Main panels: Blue and red lines indicate, respectively, ER and NR bands (mean, 10\% and 90\% contours) from our simulations of LUX's WS2013 and~WS2014-16 WIMP searches. Squares and triangles at 2~phd intervals indicate the LUX published bands. Green circles show LUX's measured events. Black contours show regions that contain 90\% of DM events. The $\log_{10}\mathrm{S2}$ scale in each panel is different. Insets: Blue and red lines show the efficiency for event detection from ERs and NRs respectively. Black triangles show the published LUX NR efficiency. }
\label{fig:S1S2}
\end{figure}

LUX~WS2014-16 was more complicated owing to the changing conditions throughout the run. We do not attempt to model the changing conditions with time or the spatially varying electric drift field. Instead we take a simplified approach and assume the average values from Ref.~\cite{Akerib:2016vxi}: $g_1=0.1~\mathrm{phd}/\gamma$, $g_2=18.9~\mathrm{phd}/e^-$, an extraction efficiency of $73\%$;  the S1 efficiency from Ref.~\cite{Bailey} and the S2 efficiency from Ref.~\cite{szydagisICHEP}; we allow events that satisfy $\mathrm{S1}_{\mathrm{raw}}>1~\mathrm{phd}$, $\mathrm{S2}_{\rm{raw}}>200~\mathrm{phd}$. We set an exclusion limit with events measured more than 1~cm from the radial fiducial volume boundary.

For LZ, we use parameter values recommended in Ref.~\cite{Akerib:2015cja}: $g_2=50~\mathrm{phd}/e^-$, an extraction efficiency of $100\%$ and allow events that satisfy $\mathrm{S2}_{\rm{raw}}>250~\mathrm{phd}$.  The S1 signal is the main determinant of the energy threshold so we show results taking the upper (lower) values of the range in Ref.~\cite{Akerib:2015cja}, namely $g_1=0.1 \,(0.05)~\mathrm{phd}/\gamma$, and allow events that satisfy~$\mathrm{S1}>2 \, (3)$~phd.

Before presenting the limits on sub-GeV DM, we demonstrate that our simulations accurately reproduce published LUX results. First, we derive the efficiency as a function of energy for ERs and NRs, shown in the insets of Fig.~\ref{fig:S1S2}, and compare against the LUX values (black triangles). Good agreement is found; within 5\% above 2~keV for both runs. Below 2~keV, we slightly underestimate the published efficiencies, reaching an underestimation of 50\% at 1.1~keV. Results for a direct comparison of the ER efficiency are not available so instead, we compare with the tritium calibration run~\cite{Akerib:2015wdi}, which had only slightly different parameters from WS2013 and WS2014-16. In the calibration run, the ER efficiency was 50\% at 1.24~keV, consistent with~1.13 and 1.25~keV that we find for WS2013 and WS2014-16, respectively.

Lastly, in Fig.~\ref{fig:S1S2}, we compare our ER and NR bands, indicated by the blue and red lines respectively, against the LUX bands, indicated by squares and triangles at 2~phd intervals (we use updated LUX WS2013 bands~\cite{ShawIDM}). We find good agreement in both the central position and the width of the bands.

\section{LUX constraints}

 \begin{figure}[!t]
\centering
\includegraphics[width=0.95\columnwidth]{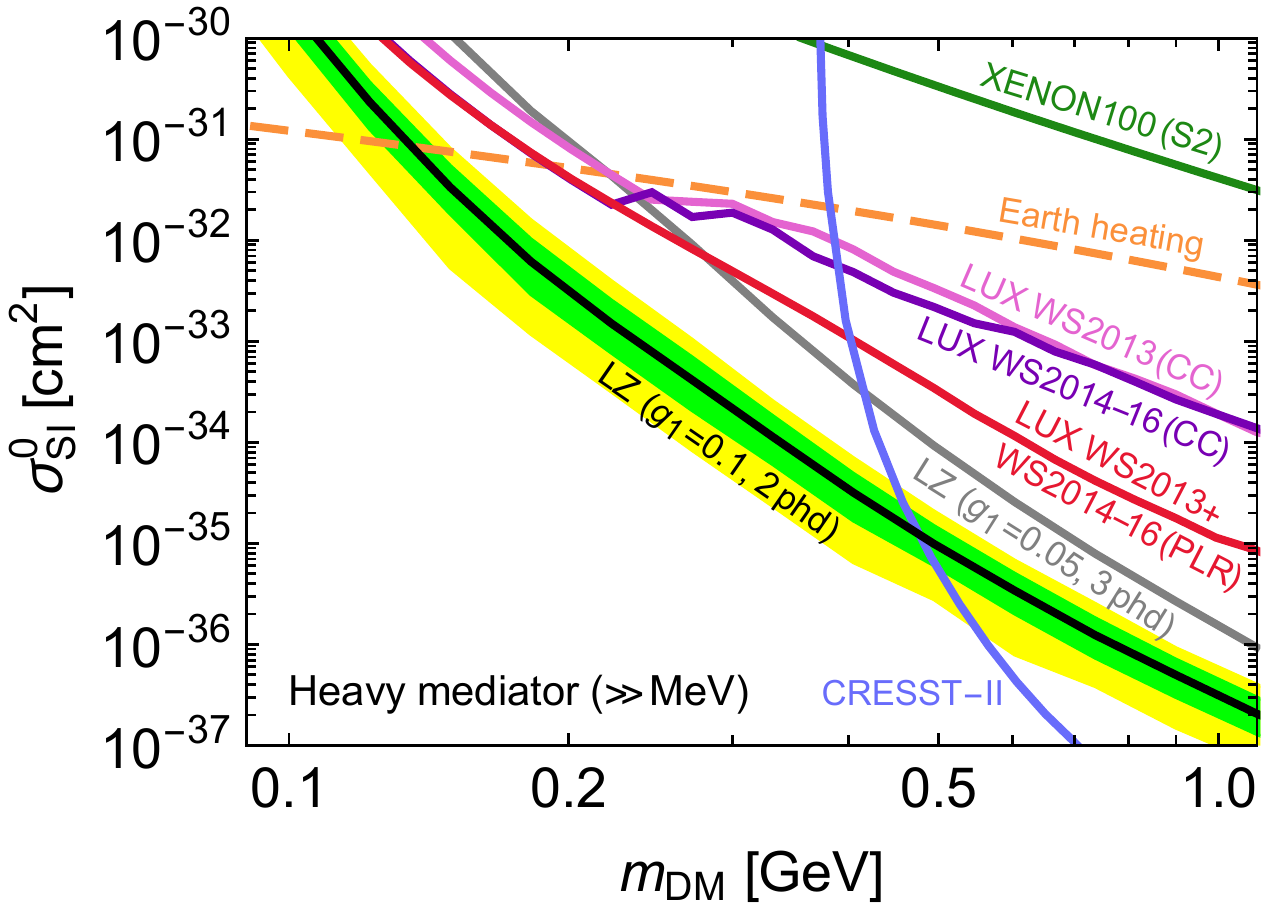}\\
\includegraphics[width=0.95\columnwidth]{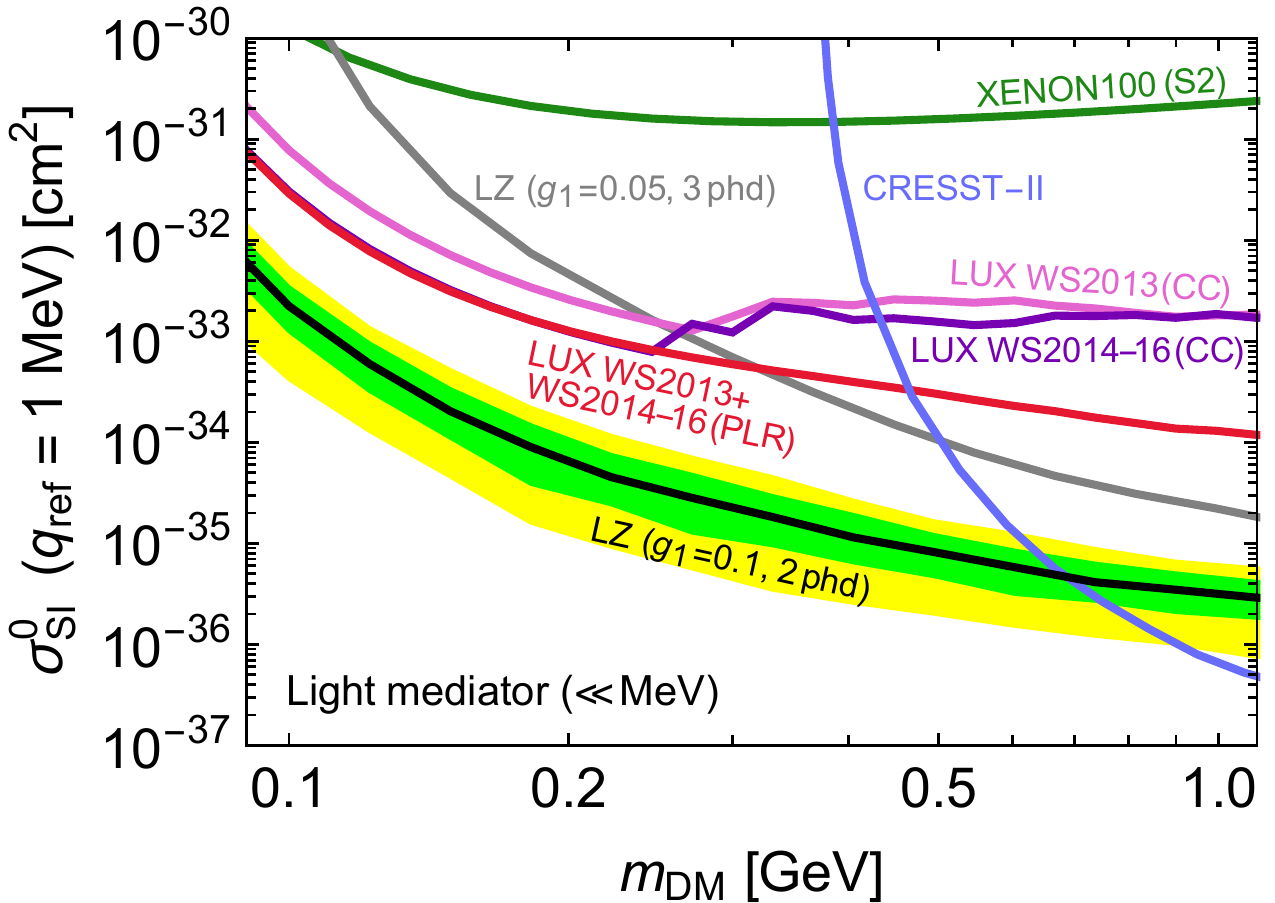}
\caption{The 90\% C.L. exclusion limits on the SI DM-nucleon cross-section. LUX WS2013 (pink line) and WS2014-16 (purple line) limits are calculated with a cut-and-count (CC) method. The combined LUX WS2013+WS2014-16 limit (red) is calculated with a profile likelihood ratio (PLR) test. XENON100 (green line) and CRESST-II (blue line) provided the most stringent exclusion limits before this work. The projected LZ sensitivity for a $3\sigma$ or greater discovery is shown in black and grey for two LZ scenarios. The upper panel includes a constraint from Earth heating by DM annihilation (orange dashed line).}
\label{fig:EXlimits}
\end{figure}

We use two methods to derive constraints on the DM-nucleon cross-section from the WS2013~\cite{Akerib:2015rjg,Akerib:2016lao} and~WS2014-16~\cite{Akerib:2016vxi} WIMP searches. 

The first method is a cut-and-count (CC) approach, the simplest and most conservative approach that treats all measured events as signal events. For each mass, we calculate the signal region that contains 90\% of the DM events that pass all cuts. The black lines in Fig.~\ref{fig:S1S2} give examples of this region for different values of~$\mDM$ in the heavy mediator limit. For $\mDM=0.1$~GeV, all of the DM events lie far below the ER band, where background events are expected to lie. For this mass, the mean~S1 signal is below the~S1 threshold so only the tail of the upward~S1 fluctuations is measured. The~S1 signal can fluctuate upwards in our simulation from the binomial modelling of the initial number of ions and excitons, the binomial modelling of the detection of photons by the photomultiplier tubes, the Gaussian resolution of the detector or finally, from the modelling of recombination fluctuations.\footnote{See Ref.~\cite{Dobi} for an extended discussion on fluctuations in LUX.} As $\mDM$ increases, more of the DM~contour overlaps the ER band because fewer upward~S1 fluctuations are probed. The signal more closely follows the ER band but there is still a small offset. For clarity, we do not show the contours for a light mediator in Fig.~\ref{fig:S1S2}. They are similar but extend to slightly smaller~S1 values e.g.~to $\mathrm{S1}= 4.4\,(5.9)$~phd for $\mDM=0.6\,(1.0)$~GeV.

We count all of the observed events within the 90\% DM contour and use Poisson statistics to set a 90\% C.L.\ exclusion limit. This is shown as the pink and purple lines in Fig.~\ref{fig:EXlimits} for the WS2013 and WS2014-16 WIMP searches respectively. The kinks around $\mDM\sim0.3$~GeV occur as measured events suddenly enter the signal region.
 
The second more powerful method to derive a constraint on the DM-nucleon cross-section uses a profile likelihood ratio (PLR) test. This takes into account the~S1 and~S2 information of each event and allows for the results from WS2013 and WS2014-16 to be combined. Unlike the CC method, the background signal must be quantified. We adopt a simple model that assumes the ER background rate is flat in energy, while ignoring subdominant contributions from neutrons and $^{8}\mathrm{B}$ neutrinos. Additionally, for the WS2013 search we include a component for the decays of $^{127}\mathrm{Xe}$, which contributed in WS2013 but not in WS2014-16. This simple model provides a good fit to LUX data~\cite{Szydagis:2016few,Akerib:2017uem} (a comparison is provided in Appendix~\ref{sec:valid}). The 90\% C.L.\ combined PLR limit is calculated following Ref.~\cite{Cowan:2010js} with an un-binned extended likelihood function~\cite{Barlow:1990vc}. We follow the safeguard method in Ref.~\cite{Priel:2016apy} to minimise the effect of background mismodelling. The amplitudes of the background components in each run are treated as a nuisance parameter. The 90\% C.L.\ limit is shown as the red line in Fig.~\ref{fig:EXlimits}. It is similar to the CC limit at low mass, where the DM signal region is far from the background region. At higher mass, the PLR limit is significantly stronger. At $\mDM=1~\mathrm{GeV}$, the limit corresponds to 5.0~signal events in WS2014-16 (for heavy and light mediators).

In Fig.~\ref{fig:EXlimits}, we also show 90\% C.L.~exclusion limits from CRESST-II~\cite{Angloher:2015ewa} and XENON100's S2-only analysis~\cite{Aprile:2016wwo}, and a constraint from Earth heating by DM annihilation~\cite{Mack:2007xj} (which does not apply if DM annihilation does not occur). For CRESST-II, we use Lise's public data~\cite{Angloher:2017zkf} and calculate a limit with the maximum gap method~\cite{Yellin:2002xd}. XENON100 observed a large number of events (13560) so we use a CC analysis with events in the range 80-1000~PE to set a limit. Our XENON100 limit is slightly weaker than in Ref.~\cite{Kouvaris:2016afs} because we adopt the $Q_y$ used in our LUX analysis, which has a cut-off at 0.19~keV. Before this work, CRESST-II and XENON100 (S2-only) provided the most stringent direct detection constraints on the DM-nucleon cross-section in this mass range. The LUX limits are significantly stronger and already reach the principal reach projected in Ref.~\cite{Kouvaris:2016afs}. This is because the displacement of signal and background regions, not previously considered, further reduces the background rate below the value considered in Ref.~\cite{Kouvaris:2016afs}.

\section{LZ sensitivity projection}

 \begin{figure}[!t]
\centering
\includegraphics[width=0.9\columnwidth]{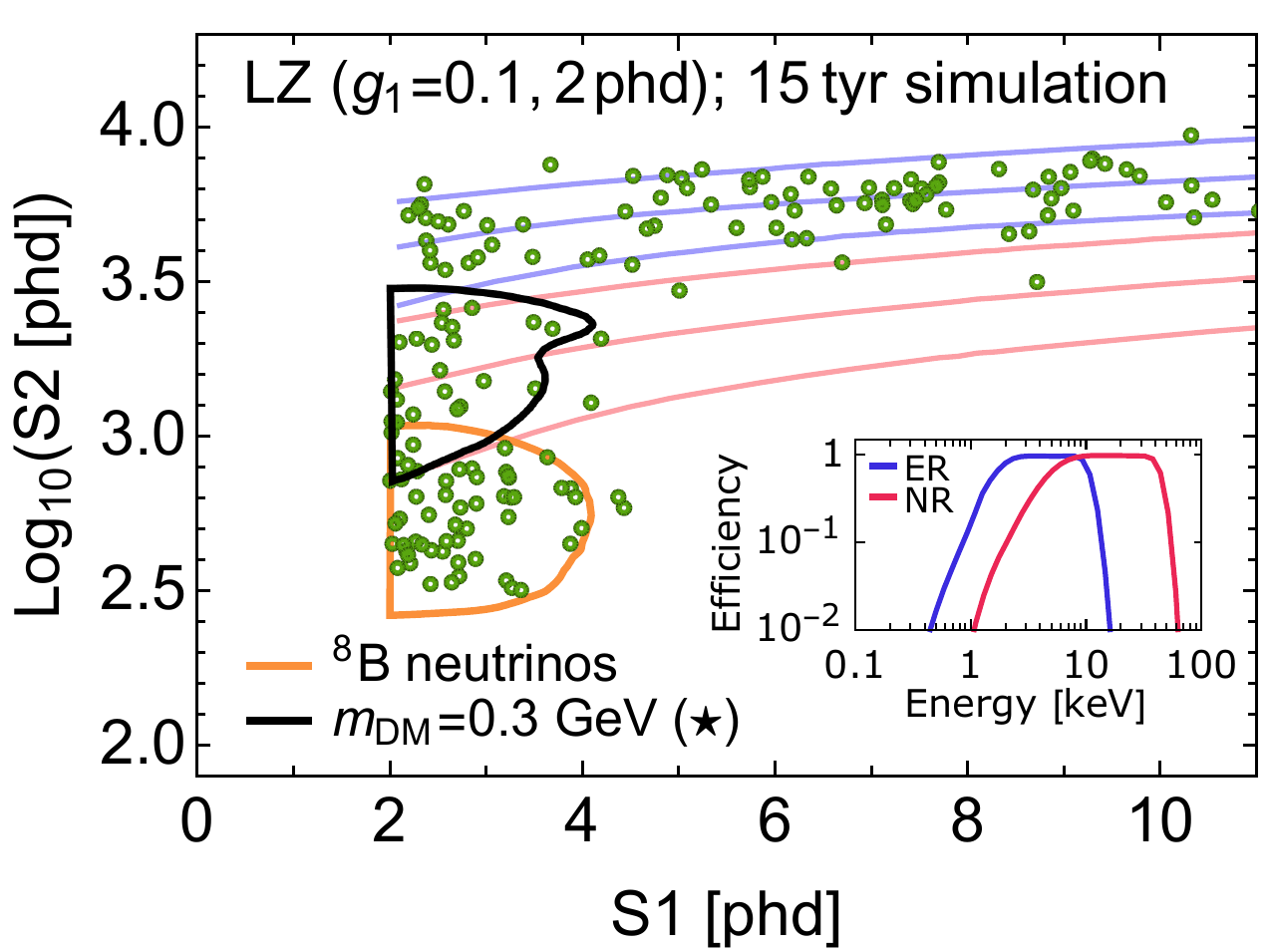}\\ \vspace{3mm}
\includegraphics[width=0.97\columnwidth]{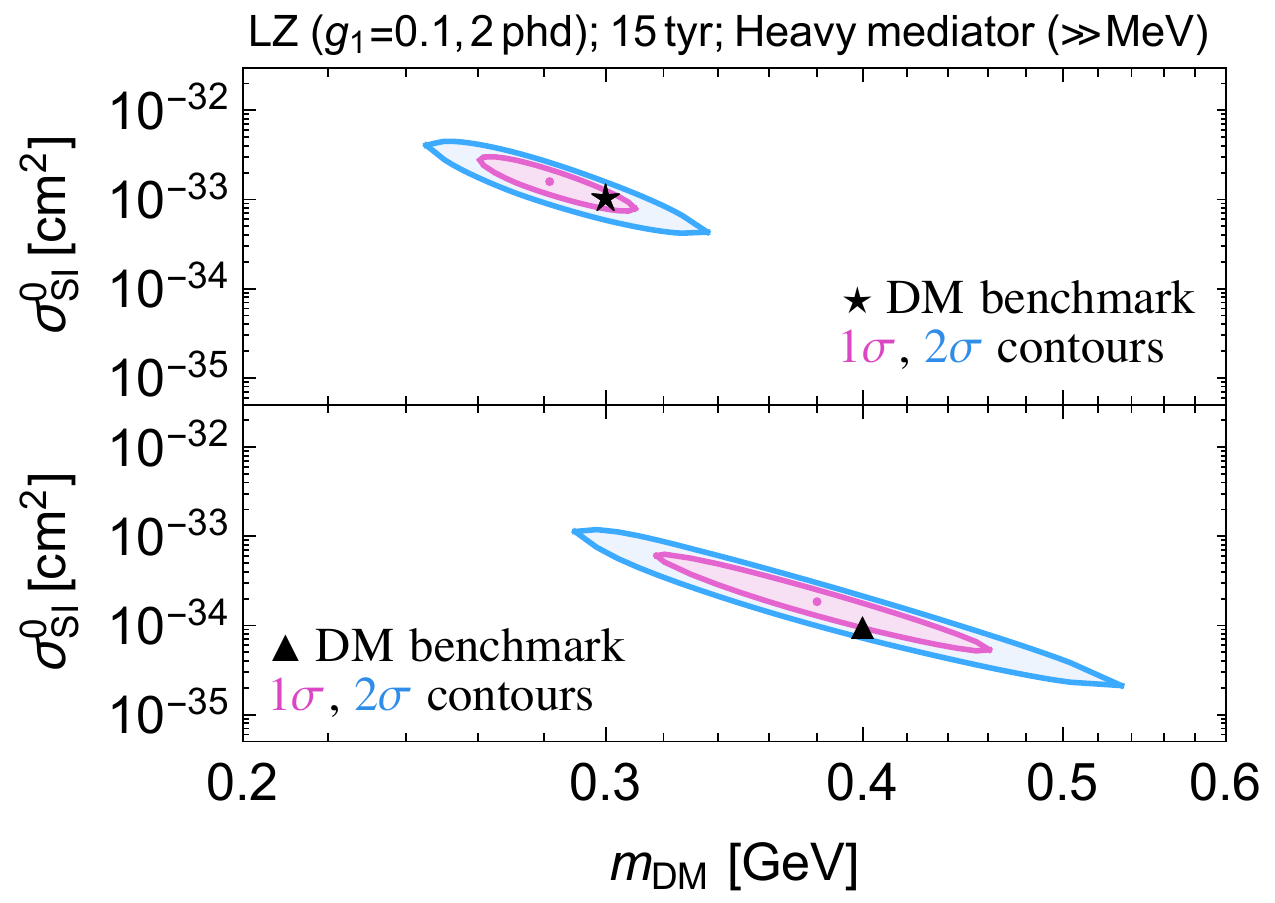}
\caption{Upper panel: Our simulation of the more sensitive LZ scenario. Blue and red lines indicate, respectively, ER and NR bands (mean, 10\% and 90\% contours). Black contours show regions that contain 90\% of DM events while the orange contour contains 90\% of $^8\mathrm{B}$ neutrino events. Green circles show simulated background and signal events from the star ($\bigstar$) DM benchmark, $\mDM=0.3$~GeV and $\sigma^0_{\rm{SI}}=10^{-33}\,\mathrm{cm}^2$. 
Lower panel: Two examples of parameter reconstruction for sub-GeV DM with LZ. The star ($\bigstar$) and triangle ($\blacktriangle$) symbols show the input mass and cross-section, corresponding to 24.6 and 21.4 expected signal events respectively.}
\label{fig:recon}
\end{figure}

The next generation of dual-phase detectors, namely LZ, XENON1T/XENONnT~\cite{Aprile:2015uzo} and PandaX-II/PandaX-xT~\cite{Cao:2014jsa}, will be bigger than LUX while having a smaller background rate. 
We focus on sensitivity projections for LZ because it has the most detailed design and performance studies~\cite{Akerib:2015cja,Mount:2017qzi}. Within LZ's fiducial volume, $^8\mathrm{B}$ (pp and $^7\mathrm{Be}$) solar neutrinos produce the dominant NR (ER) background events. Figure~\ref{fig:recon} (upper panel) shows a simulation of the events seen with the more sensitive LZ scenario that we consider ($g_1=0.1~\mathrm{phd}/\gamma$, $\mathrm{S1}\geq2~\mathrm{phd})$, together with the ER and NR efficiencies. Assuming a 5.6~tonne fiducial mass, 1000~days of data and the neutrino fluxes from Ref.~\cite{Bergstrom:2016cbh}, we find that 49.8 $^8\mathrm{B}$ events are expected. In the less sensitive scenario ($g_1=0.05~\mathrm{phd}/\gamma$, $\mathrm{S1}\geq3~\mathrm{phd})$, the ER and NR efficiencies (not shown) are shifted to higher energies. The ER (NR) efficiency is 1\% at 0.94 (2.3)~keV respectively, so that only 2.7 $^8\mathrm{B}$ events are expected.

We quantify LZ's sensitivity by calculating the median cross-section for LZ to make a discovery at $3\sigma$ (or greater) significance. This is shown for the two LZ scenarios by the black and grey lines in Fig.~\ref{fig:EXlimits}. We use a PLR test and include a 2.5\% (1\%) uncertainty on the $^8\mathrm{B}$ ($\mathrm{pp}+^7\mathrm{Be}$) flux~\cite{Bergstrom:2016cbh}. For both scenarios, these cross-sections correspond to approximately~$5\,(15)$ expected signal events at $\mDM=0.3\, (1.0)$~GeV. For the more sensitive scenario, we also show the $\pm1\sigma \,(\pm2\sigma)$ containment region in green (yellow). The more sensitive LZ scenario results in a factor 20--50 improvement compared to LUX, while the less sensitive scenario leads to only a small improvement above $\mDM=0.3$~GeV. 

For the more sensitive LZ scenario, we investigate the precision with which LZ can reconstruct the parameters of sub-GeV DM. Figure~\ref{fig:recon} (lower panel) shows examples of two reconstructions where a high-significance ($>5\sigma$) detection of DM is made. All of the signal and background events used in the reconstruction in the upper benchmark ($\bigstar$ DM benchmark) are displayed in the upper panel of Fig.~\ref{fig:recon}. For both benchmarks, the mass is reconstructed with $\simeq20\%$ accuracy.

\section{Summary}

Upcoming xenon detectors will provide opportunities to search for signals beyond the standard DM-nucleus interactions. Previous studies have investigated recoils induced from: solar neutrinos~\cite{Baudis:2013qla,Cerdeno:2016sfi}, supernova neutrinos~\cite{Chakraborty:2013zua,Davis:2016dqh,Lang:2016zhv}, nuclear~DM~\cite{Hardy:2015boa,Butcher:2016hic}, products from DM annihilation~\cite{Cherry:2015oca}; and inelastic nucleus scattering~\cite{Baudis:2013bba,McCabe:2015eia}.

We have investigated photon emission from the recoiling atom, another non-standard signal that allows dual-phase xenon detectors to probe sub-GeV DM. We have demonstrated that the LUX constraints are approximately three orders of magnitude more constraining than the S2-only limits from XENON100, and extend to smaller masses than the CRESST-II limit. In addition, a future experiment such as LZ can accurately reconstruct the parameters of sub-GeV DM since dual-phase detectors maintain the discrimination between background and signal events.

\begin{acknowledgments}
C.M. thanks Jelle Aalbers, Gianfranco Bertone, Andrew Brown, Patrick Decowski, Thomas Edwards, Chamkaur Ghag, Achim Gutlein, Federica Petricca, Ludwig Rauch, Florian Reindl, Peter Sorensen and Matthew Szydagis for discussions and correspondence. C.M. gratefully acknowledges support from the Netherlands Organisation for Scientific Research (NWO); and the Science and Technology Facilities Council (STFC) Grant ST/N004663/1.
\end{acknowledgments}

\appendix
\section{Signal models for electronic recoils \label{sec:signal}}

To aid the reproducibility of our results, we here provide the details of our signal generation model for electronic recoil (ER) events. We begin by introducing the general formalism before presenting assumptions specific to the model used in the main part of the paper. We then introduce two additional models based on the Thomas-Imel box model~\cite{Thomas:1987zz}. The material discussed below only addresses the mean signal yields. We leave the details of our model for fluctuations to the main part of the paper. 

The number of quanta $n_q$ for ER events, in terms of the number of photons $n_{\gamma}$ and electrons $n_e$, or in terms of the number of ions $n_{\rm{ion}}$ and excitons $n_{\rm{ex}}$, is
\begin{align}
\label{eq:nq}
n_q&=n_{\gamma}+n_{e}\\
&=n_{\rm{ion}}+n_{\rm{ex}}\;,
\end{align}
where $n_q = E/W$, $E$ is the energy and we take $W=13.7$~eV. 

\begin{figure}[!t]
\centering
\includegraphics[width=0.95\columnwidth]{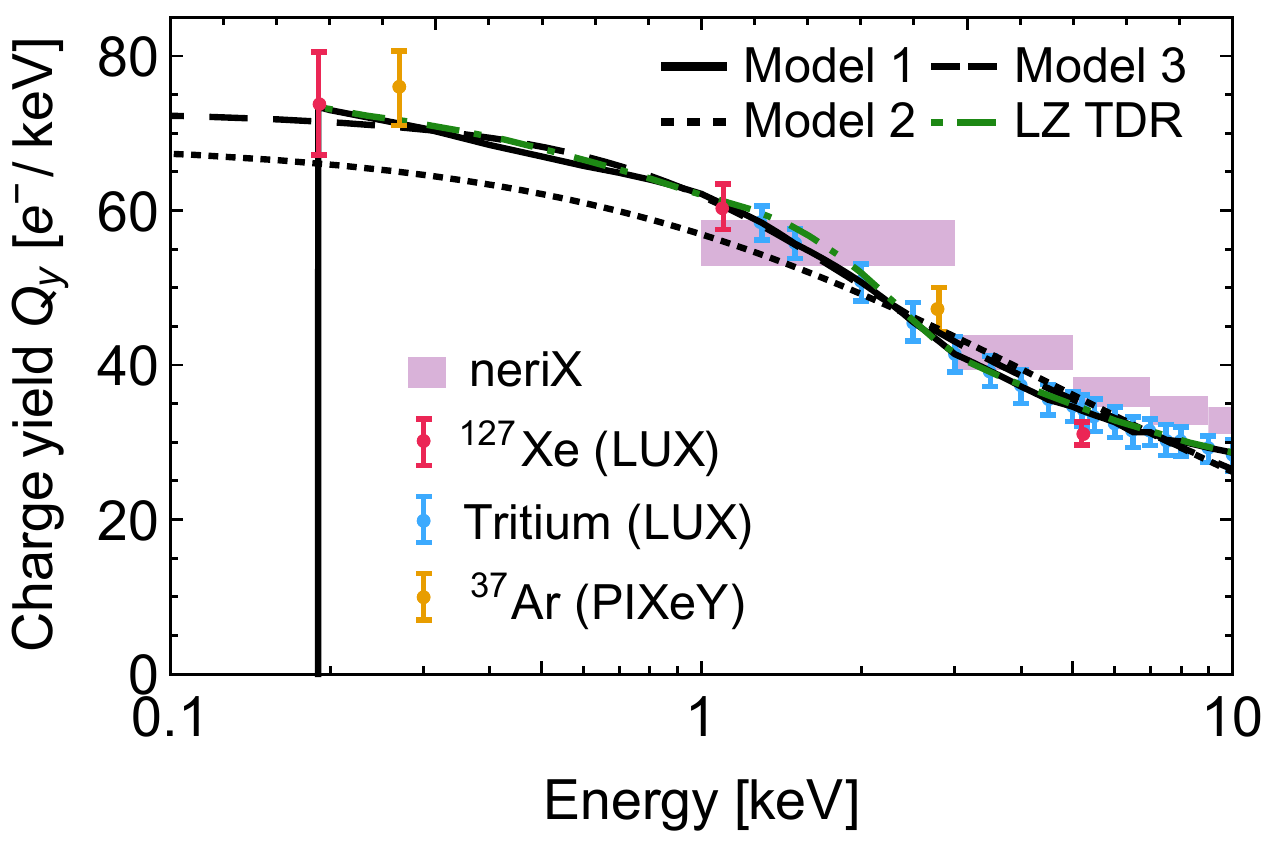}
\includegraphics[width=0.95\columnwidth]{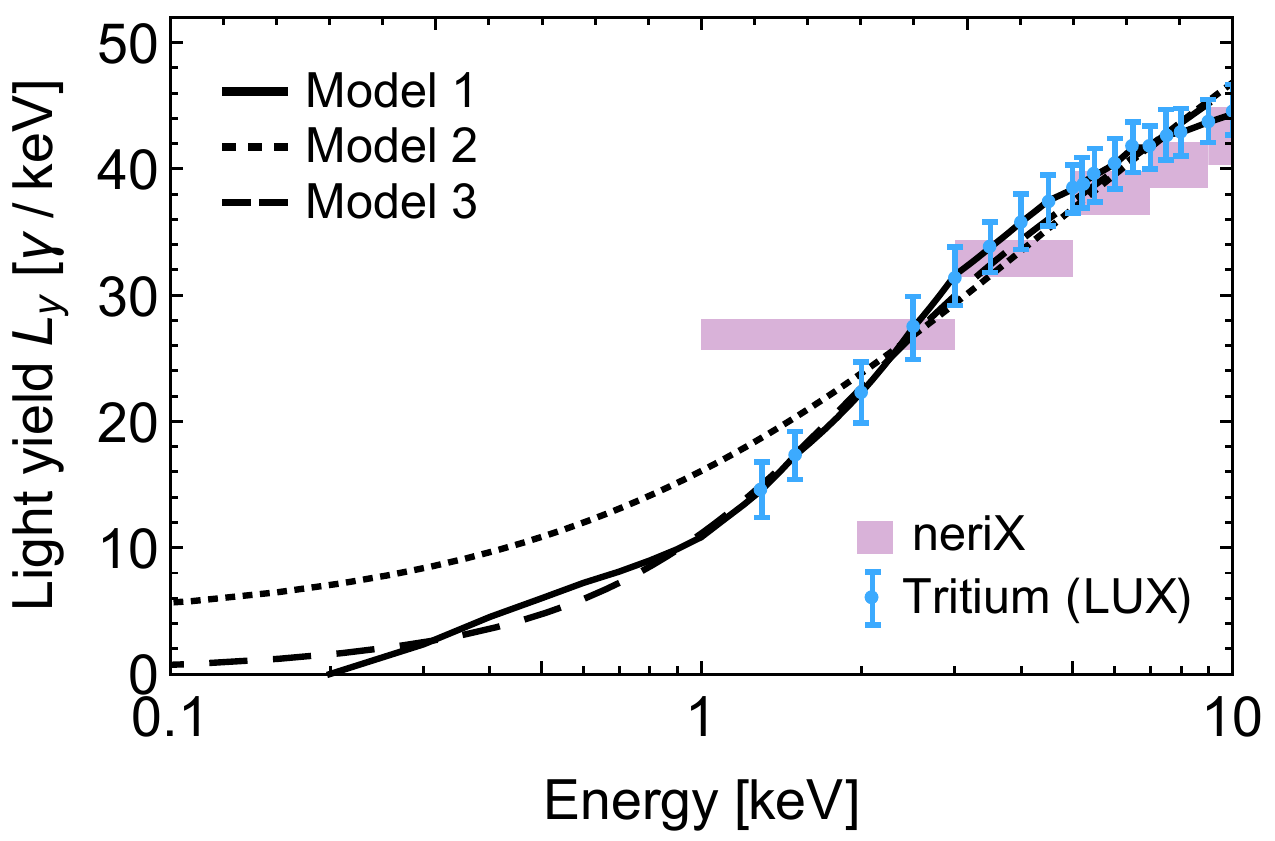}
\includegraphics[width=0.95\columnwidth]{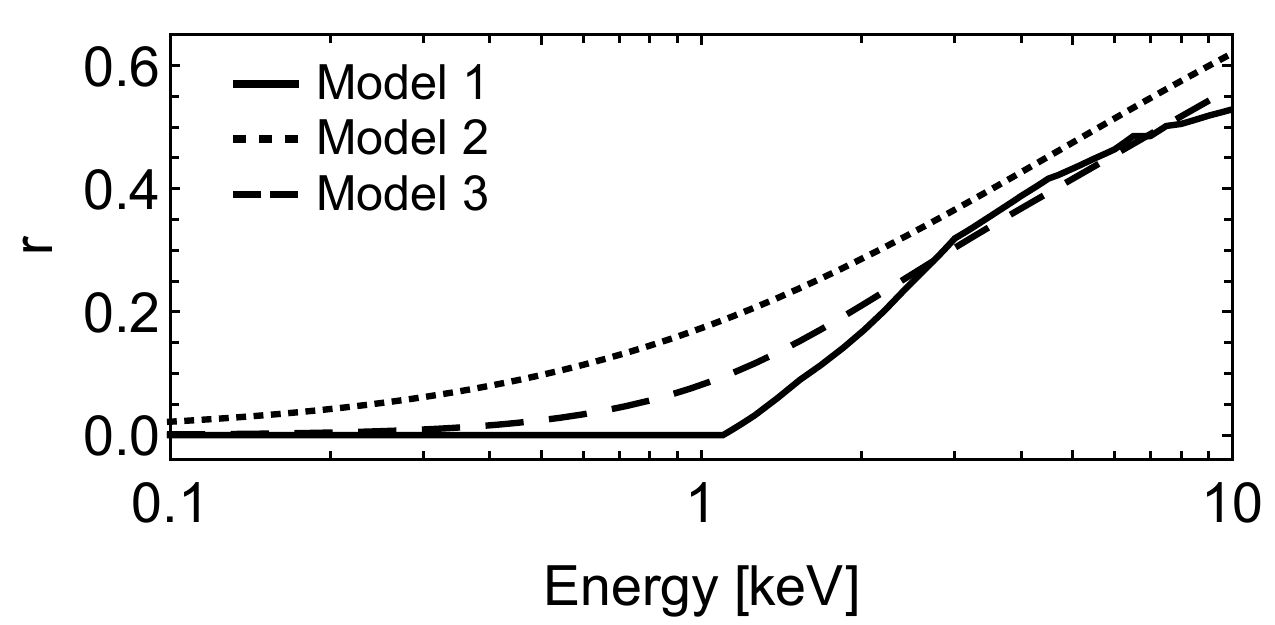}
\includegraphics[width=0.95\columnwidth]{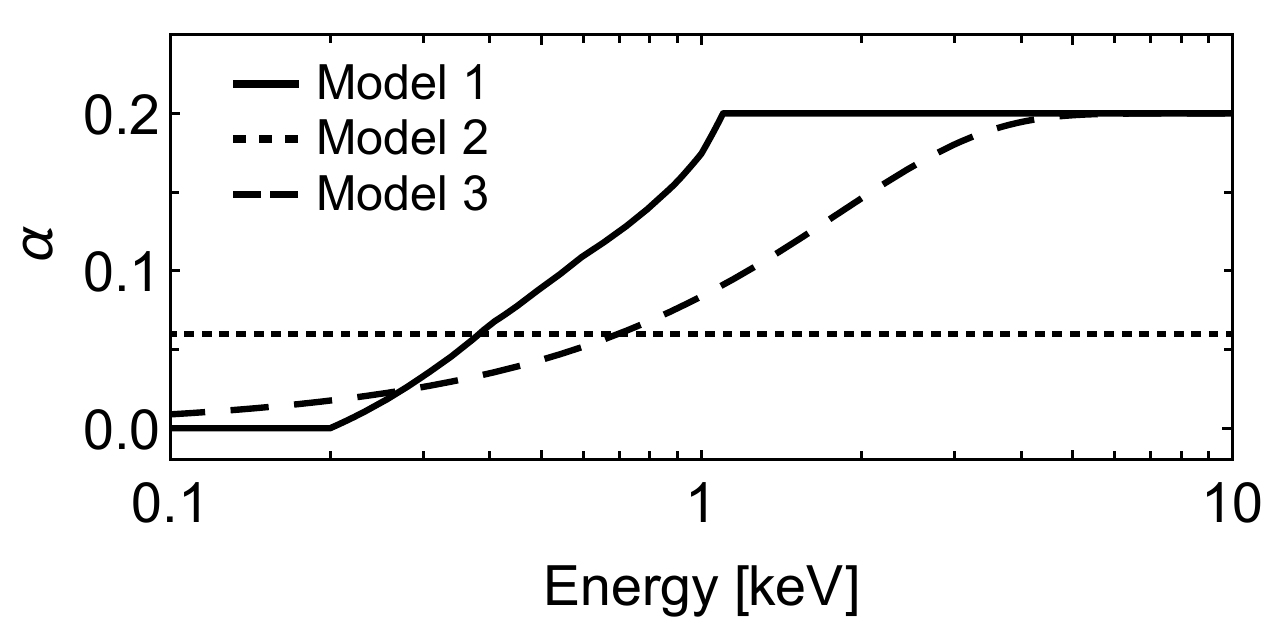}
\caption{In descending order, the panels show the electron (or charge) yield $Q_y$, the photon (or light) yield $L_y$, the recombination probability $r$ and the ratio of excitons-to-ions $\alpha$. The upper panels also include LUX's tritium and $^{127}\mathrm{Xe}$ calibration data (blue and red data points), neriX data (lilac boxes), and PIXeY's $^{37}\mathrm{Ar}$ calibration data (yellow data points). The solid, dotted and dashed black lines show the different models that we consider. Model~1 was used to generate the results in the main part of the paper.}
\label{fig:yields}
\end{figure}

The number of electrons is related to the electron yield~$Q_y$ (also referred to as the charge yield) by $n_e = Q_y E$. The starting point for our detector simulation for ERs takes $Q_y$ as input. We fit $Q_y$ to LUX's tritium calibration data~\cite{Akerib:2015wdi} above 1.3~keV, while below this, we fit to the central values from LUX's low-energy calibration with $^{127}\mathrm{Xe}$~\cite{xe127}. These data points are shown by the blue and red data points in the upper panel of Fig.~\ref{fig:yields}, respectively. In addition, the yellow data points and lilac boxes in Fig.~\ref{fig:yields} show the low-energy calibration data taken with the PIXeY and neriX xenon detectors. The PIXeY data are from the decays of $^{37}\mathrm{Ar}$~\cite{BoultonAPS,Boulton:2017hub}, while the neriX data were taken at 190~V/cm and we show the dominant systematic uncertainty~\cite{Goetzke:2016lfg}. We do not include the PIXeY or neriX data in our fits but they are consistent with the $^{127}\mathrm{Xe}$ data and tritium data.

The number of electrons is related to the number of ions through
\begin{equation}
n_e = n_{\rm{ion}} (1-r)\;,
\end{equation}
where $r$ is the fraction of ions that undergo recombination. From the sum rule in Eq.~\eqref{eq:nq}, we must have that
\begin{align}
n_{\gamma}&=n_{\rm{ex}}+r n_{\rm{ion}}\\
&= n_{\rm{ion}}(r+\alpha) \label{eq:ngamma}
\end{align}
where we have introduced the parameter $\alpha=n_{\rm{ex}}/n_{\rm{ion}}$. Values for $\alpha$ used in the literature typically fall between $0.06$ and $0.2$~\cite{Chepel:2012sj}.

\subsection{Model 1: Interpolation through the data (used in the main part of the paper) }

For the model used in the main part of the paper, $Q_y$ is determined by tracing a line (on a log axis) through the central points of LUX's tritium and $^{127}\mathrm{Xe}$ data points. This is shown by the black solid line in Fig.~\ref{fig:yields}. There is no physical basis behind this model and we include an unphysical cut-off in $Q_y$ at the energy of the lowest data point (0.19~keV). The photon yield~$L_y$ (also referred to as the light yield) is then straightforwardly determined through the relation $1/W=Q_y+L_y$ [a rearrangement of Eq.~\eqref{eq:nq}]. The solid black line in the second panel from the top of Fig.~\ref{fig:yields} shows this parameterisation of $L_y$. It passes through the central values of $L_y$ from LUX's tritium calibration data, shown by the blue data points.  The lowest $^{127}\mathrm{Xe}$ data point in the upper panel of Fig.~\ref{fig:yields} satisfies $Q_y\approx1/W$, which explains why $L_y$ tends to zero at approximately 0.19~keV.

From simple algebraic manipulation of Eqs.~\eqref{eq:nq} to~\eqref{eq:ngamma}, we can express the recombination probability $r$ as
\begin{equation}
r=\frac{L_y-\alpha Q_y}{L_y+Q_y}\;.
\end{equation}
At higher energies, we follow LUX and fix $\alpha=0.2$~\cite{Akerib:2016qlr}. Under these assumptions, $r>0$ for $E>1.1$~keV (see the solid line in the recombination panel in Fig.~\ref{fig:yields}). For energies smaller than this, we fix $r=0$ by requiring that $\alpha=L_y/Q_y$. With this approximation, $\alpha$ smoothly decreases from 0.2 to zero at approximately 0.19~keV (shown by the solid line in the bottom panel of Fig.~\ref{fig:yields}), as it must to ensure that $n_{\gamma}=0$ at approximately 0.19~keV [cf. Eq.~\eqref{eq:ngamma}].

\subsection{Model 2: Thomas-Imel box model}

The previous model was ad-hoc in that it was chosen to pass through the central value of the data points without any relation to a physically motivated model. In particular, the cut-off at approximately 0.19~keV is unphysical as it is more reasonable to expect that quanta are produced all the way to energies $\mathcal{O}(W)$. We therefore now explore the implications of a more physically motivated model of recombination: the Thomas-Imel box model~\cite{Thomas:1987zz}. In this model
\begin{equation}
r=1-\frac{\log(1+ \xi n_{\rm{ion}}) }{ \xi n_{\rm{ion}} }\;,
\end{equation}
where $\xi$ is a free parameter.

This model for $r$ allows us to rewrite the charge yield as
\begin{equation}
Q_y =\frac{1}{\xi E} \log \left(1+\frac{\xi E}{W (1+\alpha)} \right)\;.
\end{equation}
In this case, by fixing $\xi$ and $\alpha$, we determine $Q_y$, $r$ and $L_y\,(=1/W-Q_y)$. We perform a $\chi^2$ fit to the tritium and $^{127}\mathrm{Xe}$ data below 10~keV to fit $\xi$ and $\alpha$, finding $\alpha=0.06$ and $\xi=0.0065$. The resulting values of $Q_y$, $L_y$, $r$ and $\alpha$ for this model are shown by the dotted lines in Fig.~\ref{fig:yields}. This model slightly underestimates (overestimates) the $Q_y$ $(L_y)$ low-energy data points.

\subsection{Model 3: Thomas-Imel inspired model with energy-dependent coefficients}

The low-energy $^{127}\mathrm{Xe}$ and PIXeY data are consistent with $n_e \approx n_q$ (or equivalently $Q_y\approx 1/W\approx 73~e^-/\mathrm{keV}$), while in the Thomas-Imel model $Q_y\to 1/(W (1+\alpha))$ as $E \to 0$. Therefore to improve the fit with the  $^{127}\mathrm{Xe}$ and PIXeY data, we require that $\alpha \to 0$ at low energy.

\begin{figure*}[!t]
\centering
\includegraphics[width=0.66\columnwidth]{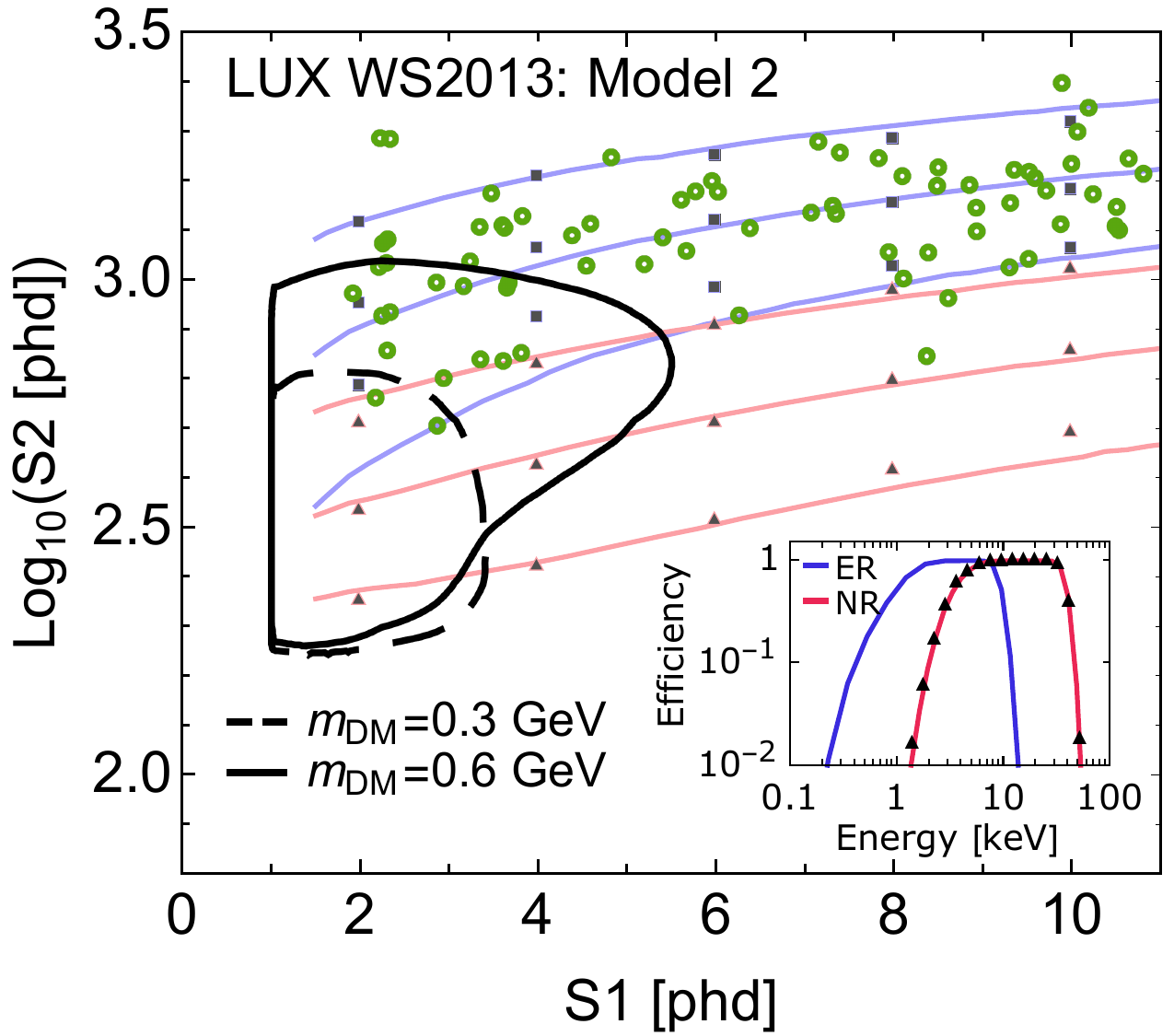}
\includegraphics[width=0.66\columnwidth]{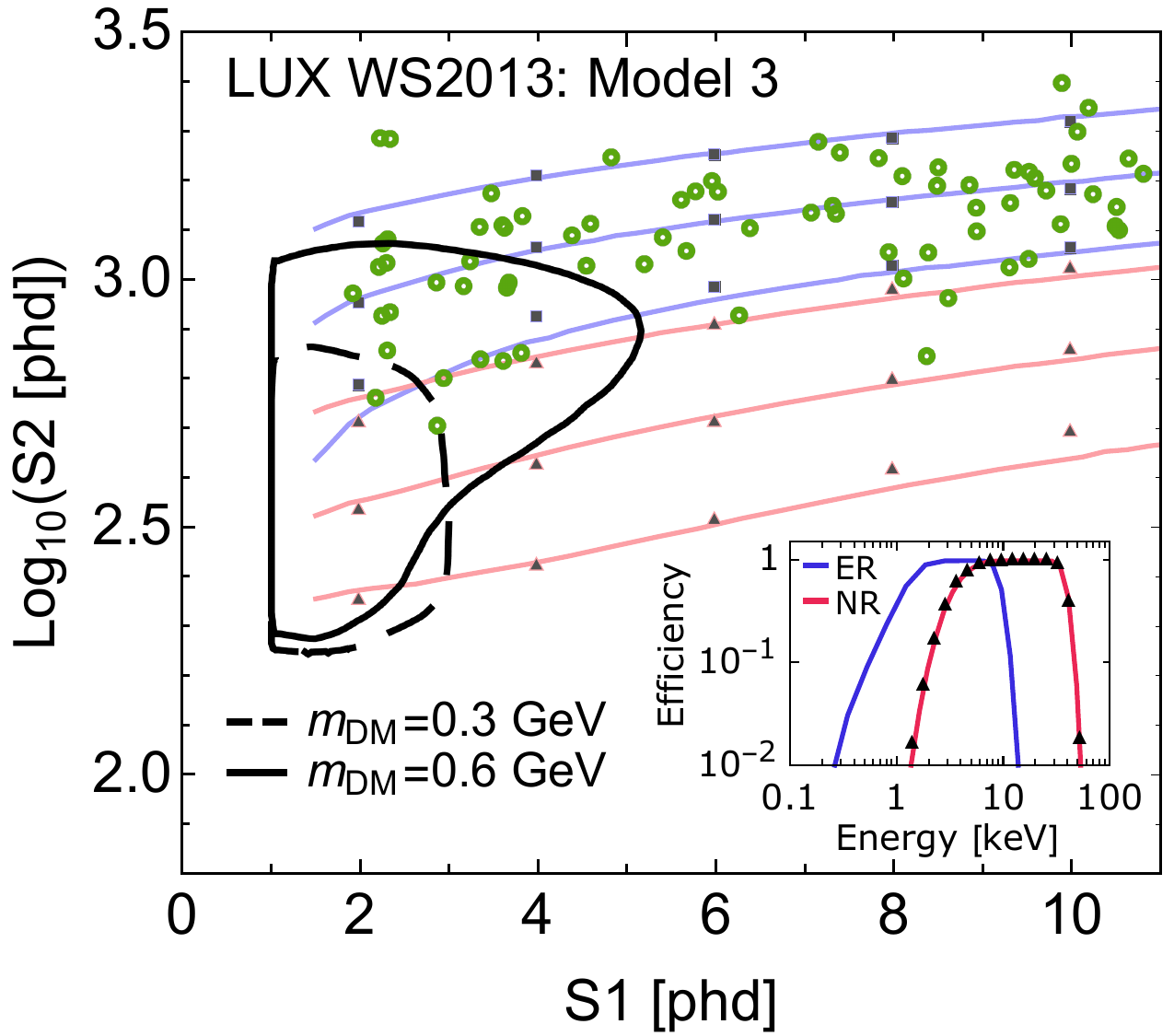}
\includegraphics[width=0.66\columnwidth]{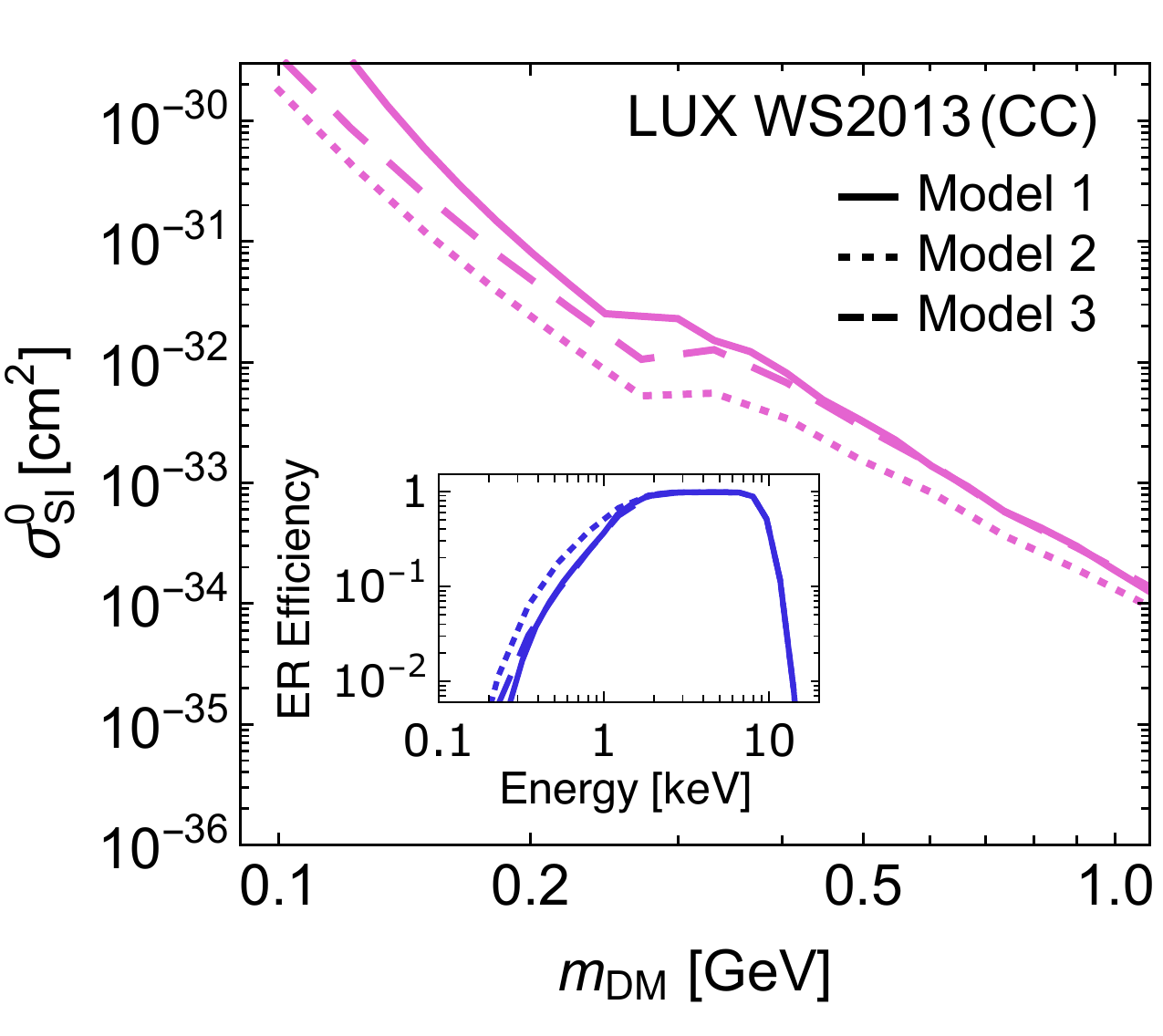}\\
\includegraphics[width=0.66\columnwidth]{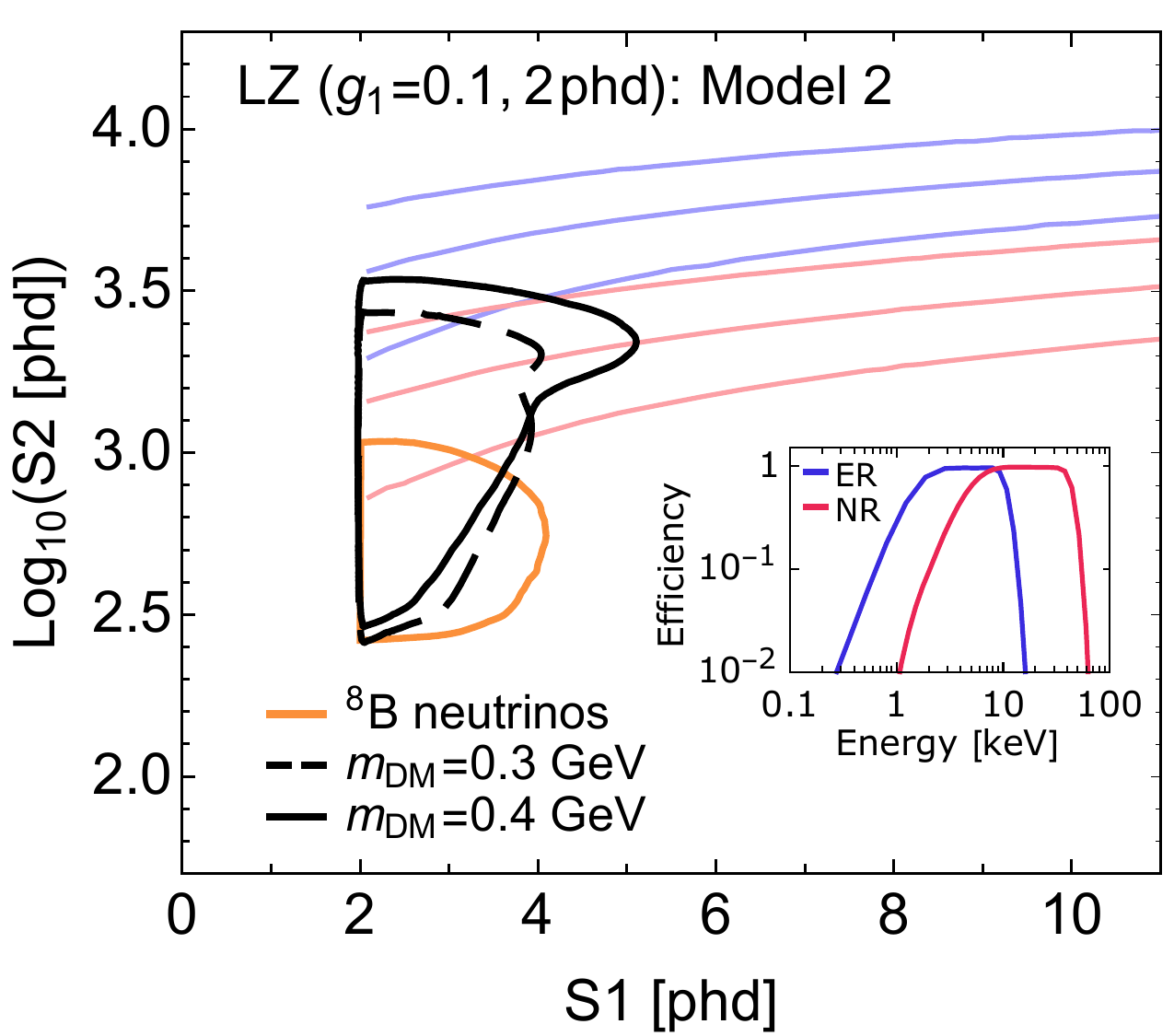}
\includegraphics[width=0.66\columnwidth]{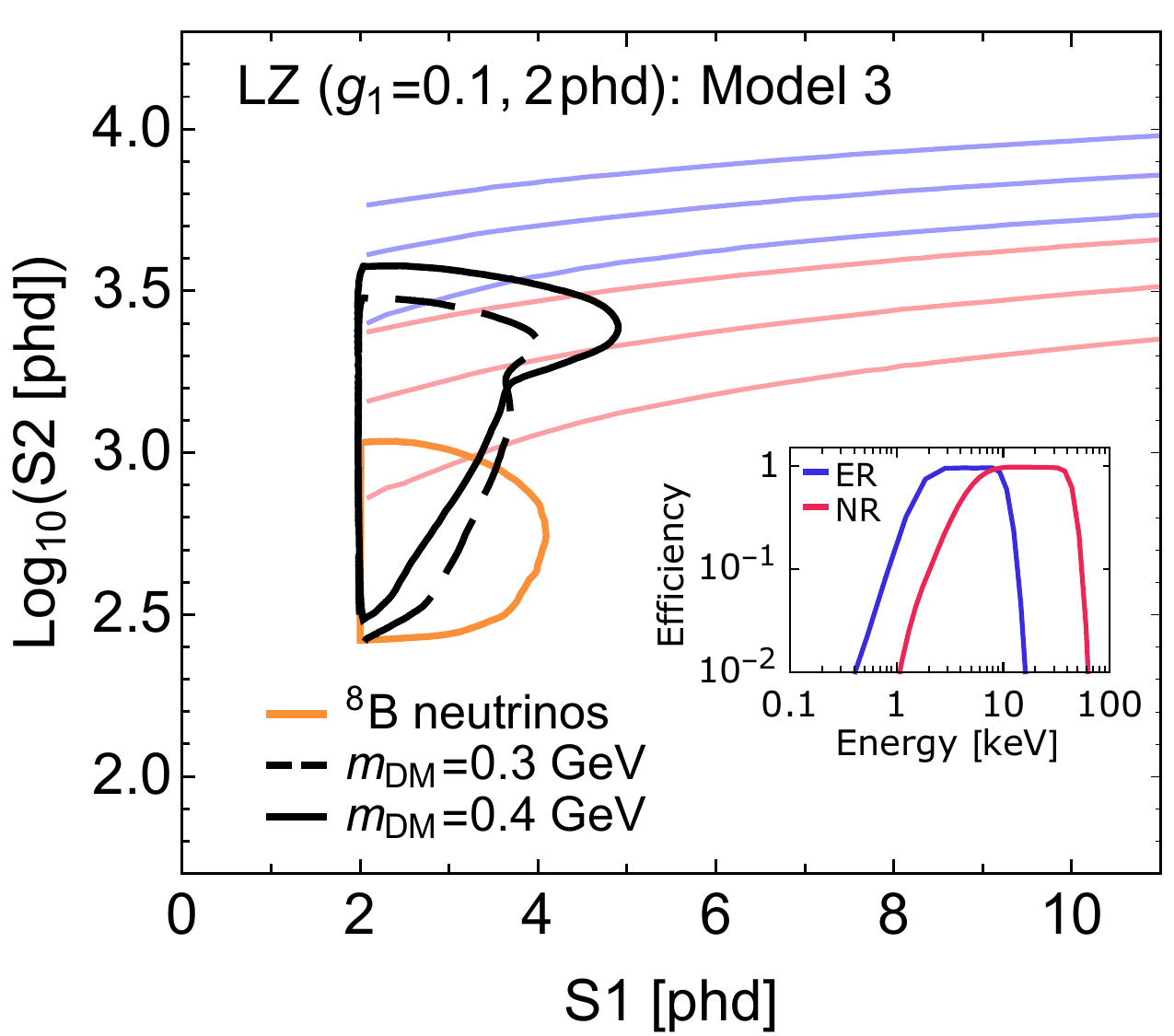}
\includegraphics[width=0.66\columnwidth]{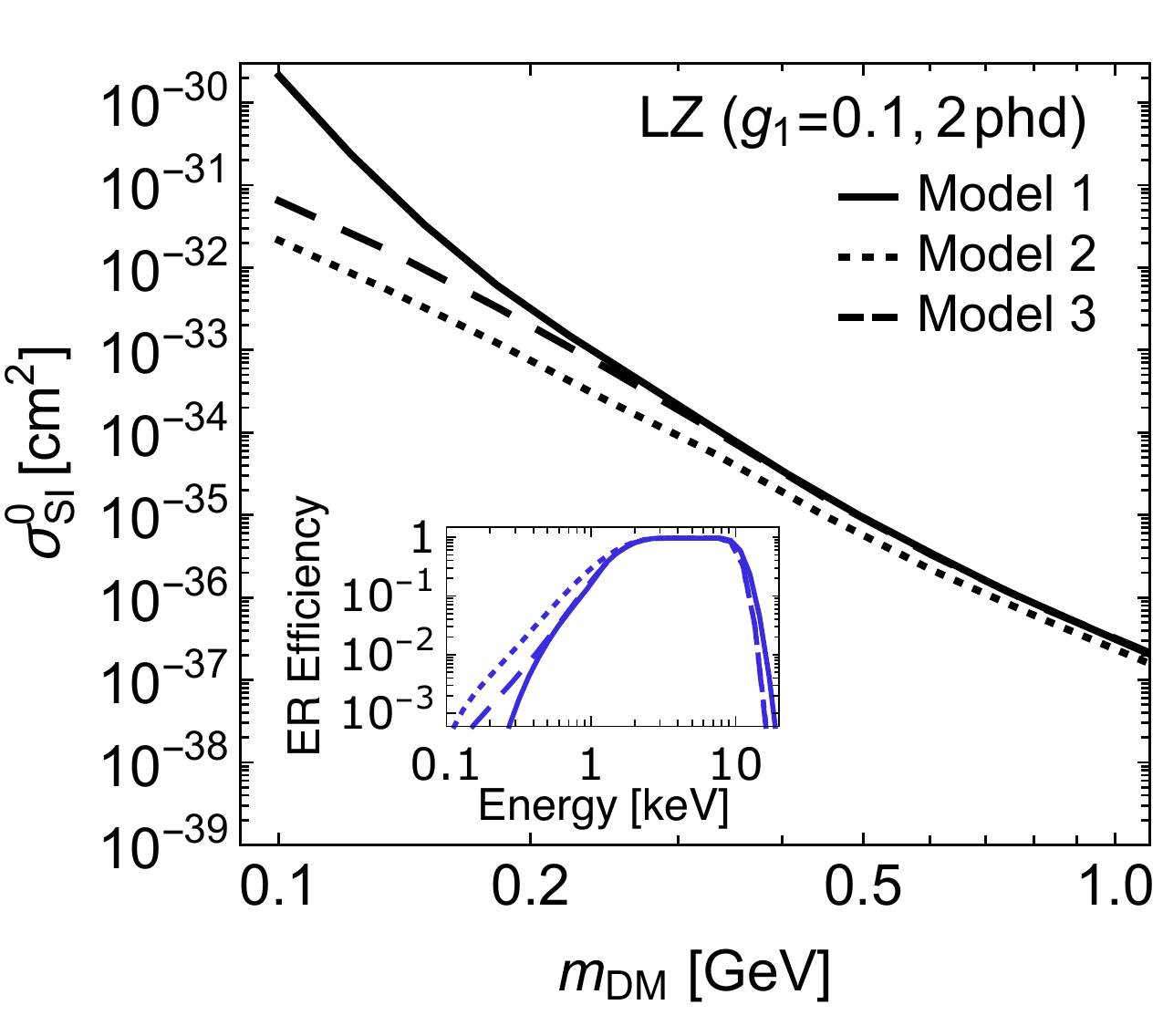}
\caption{ Left and centre panels: The distribution of background and signal regions in the~S1 vs~$\log_{10} \mathrm{S2}$ plane for our signal Models~2 and~3, for LUX~WS2013 (upper) and LZ (lower). Right panels: The upper panel shows how the LUX~WS2013 cut-and-count (CC) 90\% exclusion limits change under the different signal models, while the lower panels show the change in the median cross-section for LZ to make a discovery at $3\sigma$ (or greater) significance. The insets in the right panels show that the ER detection efficiency extends to lower energies for Models~2 and~3, which explains why the exclusion limits and discovery potential extend to smaller cross-sections in these models.}
\label{fig:limits}
\end{figure*}

We therefore modify the Thomas-Imel model to make the parameters in the Thomas-Imel model energy dependent at low energy. We define $\alpha=\alpha_0 \cdot\mathrm{erf}(\alpha_1 E)$ and $\xi = \xi_0 \cdot\mathrm{erf}(\xi_1 E)$. The justification for this parameterisation is simply to provide a way to smoothly force the parameters to zero as $E \to 0$ in order that the fit with the $^{127}\mathrm{Xe}$ and PIXeY data is improved. We perform a $\chi^2$ fit to obtain $\{\xi_0,\xi_1,\alpha_0,\alpha_1\}$. By construction, in this parameterisation $Q_y$ and $L_y$ now provide a good fit to the low-energy data points. Comparing the dashed and solid lines in Fig.~\ref{fig:yields}, we see that Model~3 is similar to Model~1 but it avoids the cutoff at 0.19~keV in~$Q_y$ and~$L_y$, and the sharp transitions in~$\alpha$ and~$r$ at 1.1~keV. 

Finally, we have also shown the $Q_y$ parameterisation adopted in the 2017 LZ technical design report (TDR)~\cite{Mount:2017qzi} by the dot-dashed green line in the upper panel of Fig.~\ref{fig:yields}. Both Model~1 and Model~3 closely resemble this parameterisation.

\section{Exclusions limits and discovery potential with different signal models \label{sec:effects}}

We now explore the impact of the different signal models on the LUX exclusion limits and the LZ discovery potential. The results for Model~1 are presented in the main part of the paper so we here focus on the results for Model~2 and Model~3.

We first focus on the exclusion limits from LUX WS2013. In the upper panels of Fig.~\ref{fig:limits}, we have recalculated the ER band (blue lines) and 90\% signal regions for two values of the dark matter mass ($\mDM$). The NR model is the same across all plots. With Model 2, the ER band extends to lower $\mathrm{S2}$ values for small $\mathrm{S1}$ values compared to Model 1 and Model 3. Meanwhile, the dark matter signal regions show only minor changes. Therefore the discrimination between signal and background is not as high in Model~2. In Models~2 and~3, the ER efficiency extends to lower energies compared to Model~1 (a direct comparison is made in the inset of the upper right panel of Fig.~\ref{fig:limits}) because in these models $L_y$ is non-zero below 0.19~keV. The result of the higher efficiency at lower energies is that the LUX limits are more stringent for Models~2 and~3 at lower values of $\mDM$ (see the limit plot in upper right panel of Fig.~\ref{fig:limits}).

We next explore the discovery potential for the more sensitive LZ scenario. In the lower panels of Fig.~\ref{fig:limits}, we have again recalculated the ER band (blue lines) and 90\% signal regions for two values of the dark matter mass. We again find that the ER band in Model~2 extends to lower $\mathrm{S2}$ values for small $\mathrm{S1}$ values compared to the other models. The signal regions are again somewhat similar, although the effect of removing the cut-off at 0.19~keV is that the DM contours extend further into the $^{8}\mathrm{B}$ region. However the signal region for these values of $\mDM$ still lies between the $^{8}\mathrm{B}$ region and the ER band (where $pp$ neutrinos and other ER background events are expected to be detected). The lower right panel of Fig.~\ref{fig:limits} shows the median cross-section for LZ to make a discovery at $3\sigma$ (or greater) significance, while the inset shows a direct comparison of the ER efficiencies. For $\mDM\sim1~\mathrm{GeV}$, the different signal models result in similar sensitivity.  The effect of removing the cut-off at 0.19~keV means that the efficiencies extend to small energies, while it allows the sensitivity of Models~2 and~3 to be enhanced with respect to Model~1 for $\mDM\sim0.1~\mathrm{GeV}$.

\section{Validation of the WS2013 background model \label{sec:valid}}

\begin{figure}[!t]
\centering
\includegraphics[width=0.99\columnwidth]{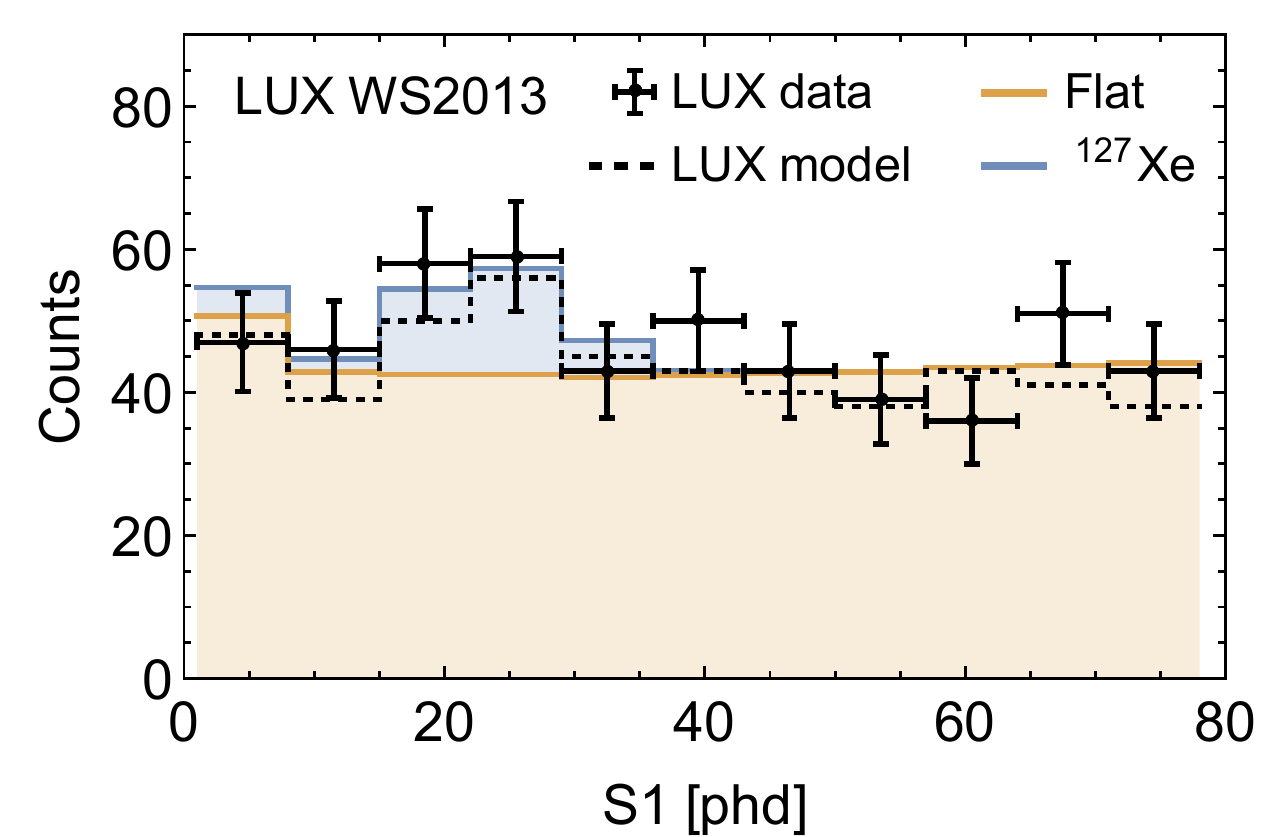}\\ \vspace{3mm}
\includegraphics[width=0.99\columnwidth]{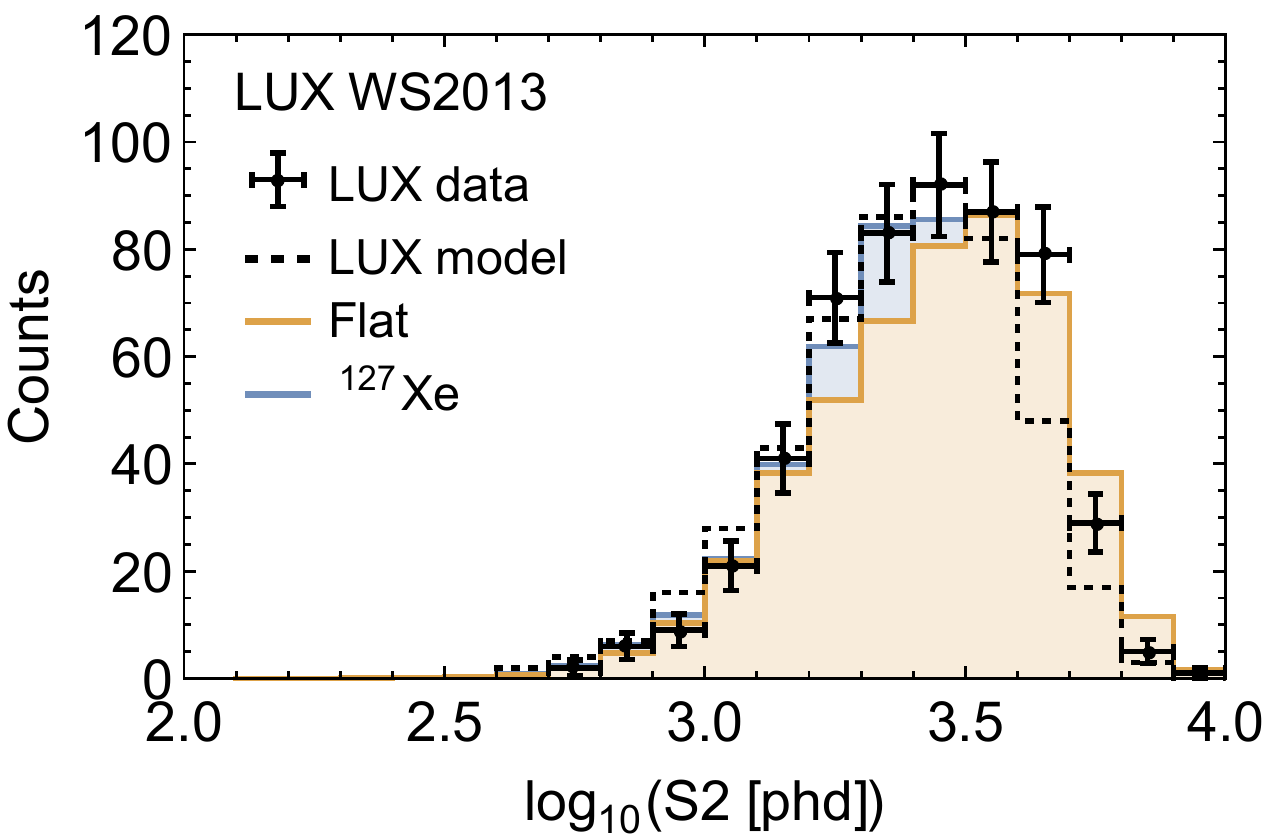}
\caption{The black data points show the background events from LUX WS2013, together with the LUX Collaboration's model (dotted black line). The yellow and blue lines show our background model, comprised of a flat component and a component from decays of $^{127}\mathrm{Xe}$, respectively. Our simple model provides a good fit to the data.}
\label{fig:validS1S2}
\end{figure} 

\begin{figure}[!t]
\centering
\vspace{5mm}
\includegraphics[width=0.99\columnwidth]{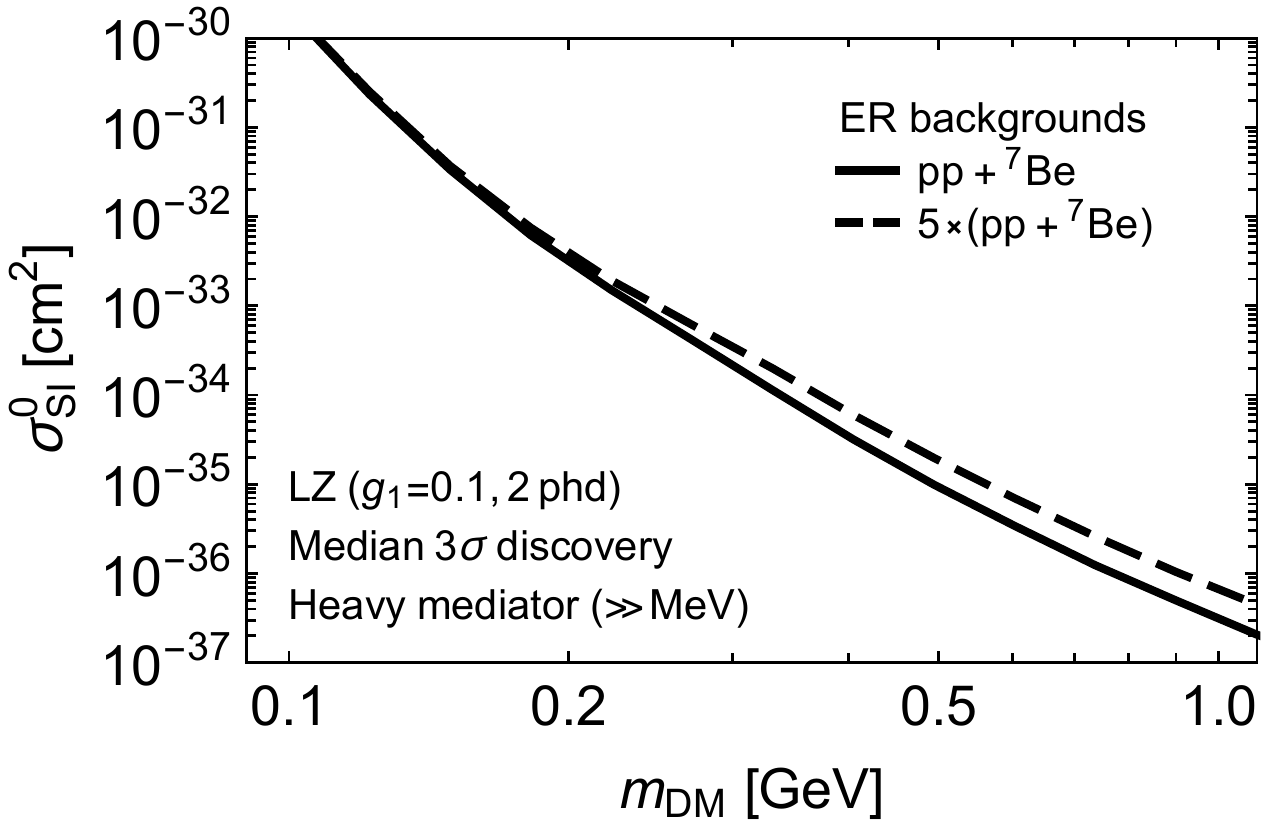}
\caption{The median cross-section for LZ to make a discovery at $3\sigma$ (or greater) significance under different assumptions for the ER background rate. The solid line assumes that the ER background is dominated by pp and $^7\mathrm{Be}$ solar neutrinos, the assumption that was made in the main part of the paper. The dashed line assumes that the ER background is five times higher than the solar neutrino rate, allowing for an additional contribution from dispersed radionuclides. The higher background rate does not change any of our conclusions.}
\label{fig:ERback}
\end{figure}

The profile likelihood ratio (PLR) test requires a background model. We here show that the simple ER background model that we have used gives a good fit to published LUX results. In particular, the background count rate and LUX's background model for WS2013 were published in their axion search paper~\cite{Akerib:2017uem}. As in this work, only events measured with a radius smaller than 18~cm were considered. The measured data points and the LUX model are shown by the black data points and dotted black line in Fig.~\ref{fig:validS1S2}, as a function of~S1 and $\log_{10} \mathrm{S2}$ in the upper and lower panels, respectively.

For our model of the WS2013 background, we assume a flat (in energy) component and a component from decays of $^{127}\mathrm{Xe}$, a cosmogenic isotope with a half-life of $\sim36$ days that decayed during WS2013 (by WS2014-16, this component had completely decayed away). Decays of $^{127}\mathrm{Xe}$ lead to ER depositions at 5.2~keV, 1.1~keV and 0.19~keV with branching ratios of 13.1\%, 2.9\% and 0.6\% respectively (we do not include the higher-energy decays that make up the remainder of the branching ratios).

The result of a fit of our model to the data is shown in Fig.~\ref{fig:validS1S2}. The yellow and blue lines show the contributions from the flat component and the $^{127}\mathrm{Xe}$ component, respectively. In both panels, we see that this simple model provides a good fit to the LUX data. It is also in good agreement with the published LUX model. The largest difference occurs in the $\log_{10} \mathrm{S2}$ comparison (lower panel) where we see that our background is slightly displaced to higher S2 values compared to the LUX model. However, our model is still in good agreement with the data points.

\section{Discovery potential with a higher background \label{sec:pp}} 

In the main part of the paper, we assumed that LZ's ER background will be dominated by pp and $^7\mathrm{Be}$ solar neutrinos. For a $5600\times1000~\mathrm{kg\,day}$ exposure, we predict 252~events in the energy range from~1.5 to 6.5~keV, which is in good agreement with the value of 255~events predicted by the LZ Collaboration (LZ also include $^{13}\mathrm{N}$ solar neutrinos which could explain the small difference)~\cite{Mount:2017qzi}. In order for the solar neutrino background to dominate, the background contribution from dispersed radionuclides (particularly radon, krypton and argon) must be sub-dominant. This was the assumption made in~Ref.~\cite{Akerib:2015cja}. However, a more recent estimate in Ref.~\cite{Mount:2017qzi} suggests that the rate from dispersed radionuclides could in fact dominate, resulting in a total of 1244~events for a $5600\times1000~\mathrm{kg\,day}$ exposure in the energy range from~1.5 to 6.5~keV.

To assess the impact of a higher background on our LZ sensitivity projection, we recalculate the median cross-section for LZ to make a discovery at $3\sigma$ (or greater) significance assuming a background rate that is five times higher than the rate from only pp and $^7\mathrm{Be}$ solar neutrinos. This results in a total of~1260~events. The solar neutrino and radionuclide energy spectra are approximately flat in energy~\cite{Schumann:2015cpa}, so a simple rescaling of the pp and $^7\mathrm{Be}$ energy spectrum is a good approximation. 

The resulting discovery cross-sections are shown in Fig.~\ref{fig:ERback} for the more sensitive LZ scenario. The black solid line shows the result in the main part of the paper, where the ER background is dominated by pp and $^7\mathrm{Be}$ solar neutrinos, while the black dashed line shows the result when the ER background is five times larger. At low masses, the dark matter signal region is far from the ER backgrounds so the result does not change. At higher masses where the impact of the ER background is most significant, the discovery cross-section is only about a factor of two higher. This is because the PLR method still has some discrimination power between signal and background owing to the slight displacement of signal and background. Thus our projections for LZ's sensitivity are robust against reasonable variations in the background rate. 

\bibliographystyle{apsrev}
\bibliography{low}

\begin{thebibliography}{91}
\expandafter\ifx\csname natexlab\endcsname\relax\def\natexlab#1{#1}\fi
\expandafter\ifx\csname bibnamefont\endcsname\relax
  \def\bibnamefont#1{#1}\fi
\expandafter\ifx\csname bibfnamefont\endcsname\relax
  \def\bibfnamefont#1{#1}\fi
\expandafter\ifx\csname citenamefont\endcsname\relax
  \def\citenamefont#1{#1}\fi
\expandafter\ifx\csname url\endcsname\relax
  \def\url#1{\texttt{#1}}\fi
\expandafter\ifx\csname urlprefix\endcsname\relax\def\urlprefix{URL }\fi
\providecommand{\bibinfo}[2]{#2}
\providecommand{\eprint}[2][]{\url{#2}}

\bibitem[{\citenamefont{Boehm and Fayet}(2004)}]{Boehm:2003hm}
\bibinfo{author}{\bibfnamefont{C.}~\bibnamefont{Boehm}} \bibnamefont{and}
  \bibinfo{author}{\bibfnamefont{P.}~\bibnamefont{Fayet}},
  \bibinfo{journal}{Nucl. Phys.} \textbf{\bibinfo{volume}{B683}},
  \bibinfo{pages}{219} (\bibinfo{year}{2004}), \eprint{hep-ph/0305261}.

\bibitem[{\citenamefont{Boehm et~al.}(2004{\natexlab{a}})\citenamefont{Boehm,
  Hooper, Silk, Casse, and Paul}}]{Boehm:2003bt}
\bibinfo{author}{\bibfnamefont{C.}~\bibnamefont{Boehm}},
  \bibinfo{author}{\bibfnamefont{D.}~\bibnamefont{Hooper}},
  \bibinfo{author}{\bibfnamefont{J.}~\bibnamefont{Silk}},
  \bibinfo{author}{\bibfnamefont{M.}~\bibnamefont{Casse}}, \bibnamefont{and}
  \bibinfo{author}{\bibfnamefont{J.}~\bibnamefont{Paul}},
  \bibinfo{journal}{Phys. Rev. Lett.} \textbf{\bibinfo{volume}{92}},
  \bibinfo{pages}{101301} (\bibinfo{year}{2004}{\natexlab{a}}),
  \eprint{astro-ph/0309686}.

\bibitem[{\citenamefont{Boehm et~al.}(2004{\natexlab{b}})\citenamefont{Boehm,
  Fayet, and Silk}}]{Boehm:2003ha}
\bibinfo{author}{\bibfnamefont{C.}~\bibnamefont{Boehm}},
  \bibinfo{author}{\bibfnamefont{P.}~\bibnamefont{Fayet}}, \bibnamefont{and}
  \bibinfo{author}{\bibfnamefont{J.}~\bibnamefont{Silk}},
  \bibinfo{journal}{Phys. Rev.} \textbf{\bibinfo{volume}{D69}},
  \bibinfo{pages}{101302} (\bibinfo{year}{2004}{\natexlab{b}}),
  \eprint{hep-ph/0311143}.

\bibitem[{\citenamefont{Pospelov
  et~al.}(2008{\natexlab{a}})\citenamefont{Pospelov, Ritz, and
  Voloshin}}]{Pospelov:2007mp}
\bibinfo{author}{\bibfnamefont{M.}~\bibnamefont{Pospelov}},
  \bibinfo{author}{\bibfnamefont{A.}~\bibnamefont{Ritz}}, \bibnamefont{and}
  \bibinfo{author}{\bibfnamefont{M.~B.} \bibnamefont{Voloshin}},
  \bibinfo{journal}{Phys. Lett.} \textbf{\bibinfo{volume}{B662}},
  \bibinfo{pages}{53} (\bibinfo{year}{2008}{\natexlab{a}}), \eprint{0711.4866}.

\bibitem[{\citenamefont{Hall et~al.}(2010)\citenamefont{Hall, Jedamzik,
  March-Russell, and West}}]{Hall:2009bx}
\bibinfo{author}{\bibfnamefont{L.~J.} \bibnamefont{Hall}},
  \bibinfo{author}{\bibfnamefont{K.}~\bibnamefont{Jedamzik}},
  \bibinfo{author}{\bibfnamefont{J.}~\bibnamefont{March-Russell}},
  \bibnamefont{and} \bibinfo{author}{\bibfnamefont{S.~M.} \bibnamefont{West}},
  \bibinfo{journal}{JHEP} \textbf{\bibinfo{volume}{03}}, \bibinfo{pages}{080}
  (\bibinfo{year}{2010}), \eprint{0911.1120}.

\bibitem[{\citenamefont{Hochberg et~al.}(2014)\citenamefont{Hochberg, Kuflik,
  Volansky, and Wacker}}]{Hochberg:2014dra}
\bibinfo{author}{\bibfnamefont{Y.}~\bibnamefont{Hochberg}},
  \bibinfo{author}{\bibfnamefont{E.}~\bibnamefont{Kuflik}},
  \bibinfo{author}{\bibfnamefont{T.}~\bibnamefont{Volansky}}, \bibnamefont{and}
  \bibinfo{author}{\bibfnamefont{J.~G.} \bibnamefont{Wacker}},
  \bibinfo{journal}{Phys. Rev. Lett.} \textbf{\bibinfo{volume}{113}},
  \bibinfo{pages}{171301} (\bibinfo{year}{2014}), \eprint{1402.5143}.

\bibitem[{\citenamefont{D'Agnolo and Ruderman}(2015)}]{DAgnolo:2015ujb}
\bibinfo{author}{\bibfnamefont{R.~T.} \bibnamefont{D'Agnolo}} \bibnamefont{and}
  \bibinfo{author}{\bibfnamefont{J.~T.} \bibnamefont{Ruderman}},
  \bibinfo{journal}{Phys. Rev. Lett.} \textbf{\bibinfo{volume}{115}},
  \bibinfo{pages}{061301} (\bibinfo{year}{2015}), \eprint{1505.07107}.

\bibitem[{\citenamefont{Pappadopulo et~al.}(2016)\citenamefont{Pappadopulo,
  Ruderman, and Trevisan}}]{Pappadopulo:2016pkp}
\bibinfo{author}{\bibfnamefont{D.}~\bibnamefont{Pappadopulo}},
  \bibinfo{author}{\bibfnamefont{J.~T.} \bibnamefont{Ruderman}},
  \bibnamefont{and} \bibinfo{author}{\bibfnamefont{G.}~\bibnamefont{Trevisan}},
  \bibinfo{journal}{Phys. Rev.} \textbf{\bibinfo{volume}{D94}},
  \bibinfo{pages}{035005} (\bibinfo{year}{2016}), \eprint{1602.04219}.

\bibitem[{\citenamefont{D'Agnolo et~al.}(2017)\citenamefont{D'Agnolo,
  Pappadopulo, and Ruderman}}]{DAgnolo:2017dbv}
\bibinfo{author}{\bibfnamefont{R.~T.} \bibnamefont{D'Agnolo}},
  \bibinfo{author}{\bibfnamefont{D.}~\bibnamefont{Pappadopulo}},
  \bibnamefont{and} \bibinfo{author}{\bibfnamefont{J.~T.}
  \bibnamefont{Ruderman}} (\bibinfo{year}{2017}), \eprint{1705.08450}.

\bibitem[{\citenamefont{Essig et~al.}(2016)\citenamefont{Essig,
  Fernandez-Serra, Mardon, Soto, Volansky, and Yu}}]{Essig:2015cda}
\bibinfo{author}{\bibfnamefont{R.}~\bibnamefont{Essig}},
  \bibinfo{author}{\bibfnamefont{M.}~\bibnamefont{Fernandez-Serra}},
  \bibinfo{author}{\bibfnamefont{J.}~\bibnamefont{Mardon}},
  \bibinfo{author}{\bibfnamefont{A.}~\bibnamefont{Soto}},
  \bibinfo{author}{\bibfnamefont{T.}~\bibnamefont{Volansky}}, \bibnamefont{and}
  \bibinfo{author}{\bibfnamefont{T.-T.} \bibnamefont{Yu}},
  \bibinfo{journal}{JHEP} \textbf{\bibinfo{volume}{05}}, \bibinfo{pages}{046}
  (\bibinfo{year}{2016}), \eprint{1509.01598}.

\bibitem[{\citenamefont{Hochberg
  et~al.}(2016{\natexlab{a}})\citenamefont{Hochberg, Pyle, Zhao, and
  Zurek}}]{Hochberg:2015fth}
\bibinfo{author}{\bibfnamefont{Y.}~\bibnamefont{Hochberg}},
  \bibinfo{author}{\bibfnamefont{M.}~\bibnamefont{Pyle}},
  \bibinfo{author}{\bibfnamefont{Y.}~\bibnamefont{Zhao}}, \bibnamefont{and}
  \bibinfo{author}{\bibfnamefont{K.~M.} \bibnamefont{Zurek}},
  \bibinfo{journal}{JHEP} \textbf{\bibinfo{volume}{08}}, \bibinfo{pages}{057}
  (\bibinfo{year}{2016}{\natexlab{a}}), \eprint{1512.04533}.

\bibitem[{\citenamefont{Hochberg
  et~al.}(2017{\natexlab{a}})\citenamefont{Hochberg, Kahn, Lisanti, Tully, and
  Zurek}}]{Hochberg:2016ntt}
\bibinfo{author}{\bibfnamefont{Y.}~\bibnamefont{Hochberg}},
  \bibinfo{author}{\bibfnamefont{Y.}~\bibnamefont{Kahn}},
  \bibinfo{author}{\bibfnamefont{M.}~\bibnamefont{Lisanti}},
  \bibinfo{author}{\bibfnamefont{C.~G.} \bibnamefont{Tully}}, \bibnamefont{and}
  \bibinfo{author}{\bibfnamefont{K.~M.} \bibnamefont{Zurek}},
  \bibinfo{journal}{Phys. Lett.} \textbf{\bibinfo{volume}{B772}},
  \bibinfo{pages}{239} (\bibinfo{year}{2017}{\natexlab{a}}),
  \eprint{1606.08849}.

\bibitem[{\citenamefont{Derenzo et~al.}(2017)\citenamefont{Derenzo, Essig,
  Massari, Soto, and Yu}}]{Derenzo:2016fse}
\bibinfo{author}{\bibfnamefont{S.}~\bibnamefont{Derenzo}},
  \bibinfo{author}{\bibfnamefont{R.}~\bibnamefont{Essig}},
  \bibinfo{author}{\bibfnamefont{A.}~\bibnamefont{Massari}},
  \bibinfo{author}{\bibfnamefont{A.}~\bibnamefont{Soto}}, \bibnamefont{and}
  \bibinfo{author}{\bibfnamefont{T.-T.} \bibnamefont{Yu}},
  \bibinfo{journal}{Phys. Rev.} \textbf{\bibinfo{volume}{D96}},
  \bibinfo{pages}{016026} (\bibinfo{year}{2017}), \eprint{1607.01009}.

\bibitem[{\citenamefont{Hochberg
  et~al.}(2016{\natexlab{b}})\citenamefont{Hochberg, Lin, and
  Zurek}}]{Hochberg:2016ajh}
\bibinfo{author}{\bibfnamefont{Y.}~\bibnamefont{Hochberg}},
  \bibinfo{author}{\bibfnamefont{T.}~\bibnamefont{Lin}}, \bibnamefont{and}
  \bibinfo{author}{\bibfnamefont{K.~M.} \bibnamefont{Zurek}},
  \bibinfo{journal}{Phys. Rev.} \textbf{\bibinfo{volume}{D94}},
  \bibinfo{pages}{015019} (\bibinfo{year}{2016}{\natexlab{b}}),
  \eprint{1604.06800}.

\bibitem[{\citenamefont{Hochberg
  et~al.}(2017{\natexlab{b}})\citenamefont{Hochberg, Lin, and
  Zurek}}]{Hochberg:2016sqx}
\bibinfo{author}{\bibfnamefont{Y.}~\bibnamefont{Hochberg}},
  \bibinfo{author}{\bibfnamefont{T.}~\bibnamefont{Lin}}, \bibnamefont{and}
  \bibinfo{author}{\bibfnamefont{K.~M.} \bibnamefont{Zurek}},
  \bibinfo{journal}{Phys. Rev.} \textbf{\bibinfo{volume}{D95}},
  \bibinfo{pages}{023013} (\bibinfo{year}{2017}{\natexlab{b}}),
  \eprint{1608.01994}.

\bibitem[{\citenamefont{Bunting et~al.}(2017)\citenamefont{Bunting, Gratta,
  Melia, and Rajendran}}]{Bunting:2017net}
\bibinfo{author}{\bibfnamefont{P.~C.} \bibnamefont{Bunting}},
  \bibinfo{author}{\bibfnamefont{G.}~\bibnamefont{Gratta}},
  \bibinfo{author}{\bibfnamefont{T.}~\bibnamefont{Melia}}, \bibnamefont{and}
  \bibinfo{author}{\bibfnamefont{S.}~\bibnamefont{Rajendran}},
  \bibinfo{journal}{Phys. Rev.} \textbf{\bibinfo{volume}{D95}},
  \bibinfo{pages}{095001} (\bibinfo{year}{2017}), \eprint{1701.06566}.

\bibitem[{\citenamefont{Guo and McKinsey}(2013)}]{Guo:2013dt}
\bibinfo{author}{\bibfnamefont{W.}~\bibnamefont{Guo}} \bibnamefont{and}
  \bibinfo{author}{\bibfnamefont{D.~N.} \bibnamefont{McKinsey}},
  \bibinfo{journal}{Phys. Rev.} \textbf{\bibinfo{volume}{D87}},
  \bibinfo{pages}{115001} (\bibinfo{year}{2013}), \eprint{1302.0534}.

\bibitem[{\citenamefont{Schutz and Zurek}(2016)}]{Schutz:2016tid}
\bibinfo{author}{\bibfnamefont{K.}~\bibnamefont{Schutz}} \bibnamefont{and}
  \bibinfo{author}{\bibfnamefont{K.~M.} \bibnamefont{Zurek}},
  \bibinfo{journal}{Phys. Rev. Lett.} \textbf{\bibinfo{volume}{117}},
  \bibinfo{pages}{121302} (\bibinfo{year}{2016}), \eprint{1604.08206}.

\bibitem[{\citenamefont{Knapen et~al.}(2017)\citenamefont{Knapen, Lin, and
  Zurek}}]{Knapen:2016cue}
\bibinfo{author}{\bibfnamefont{S.}~\bibnamefont{Knapen}},
  \bibinfo{author}{\bibfnamefont{T.}~\bibnamefont{Lin}}, \bibnamefont{and}
  \bibinfo{author}{\bibfnamefont{K.~M.} \bibnamefont{Zurek}},
  \bibinfo{journal}{Phys. Rev.} \textbf{\bibinfo{volume}{D95}},
  \bibinfo{pages}{056019} (\bibinfo{year}{2017}), \eprint{1611.06228}.

\bibitem[{\citenamefont{Essig et~al.}(2017)\citenamefont{Essig, Mardon, Slone,
  and Volansky}}]{Essig:2016crl}
\bibinfo{author}{\bibfnamefont{R.}~\bibnamefont{Essig}},
  \bibinfo{author}{\bibfnamefont{J.}~\bibnamefont{Mardon}},
  \bibinfo{author}{\bibfnamefont{O.}~\bibnamefont{Slone}}, \bibnamefont{and}
  \bibinfo{author}{\bibfnamefont{T.}~\bibnamefont{Volansky}},
  \bibinfo{journal}{Phys. Rev.} \textbf{\bibinfo{volume}{D95}},
  \bibinfo{pages}{056011} (\bibinfo{year}{2017}), \eprint{1608.02940}.

\bibitem[{\citenamefont{Pospelov
  et~al.}(2008{\natexlab{b}})\citenamefont{Pospelov, Ritz, and
  Voloshin}}]{Pospelov:2008jk}
\bibinfo{author}{\bibfnamefont{M.}~\bibnamefont{Pospelov}},
  \bibinfo{author}{\bibfnamefont{A.}~\bibnamefont{Ritz}}, \bibnamefont{and}
  \bibinfo{author}{\bibfnamefont{M.~B.} \bibnamefont{Voloshin}},
  \bibinfo{journal}{Phys. Rev.} \textbf{\bibinfo{volume}{D78}},
  \bibinfo{pages}{115012} (\bibinfo{year}{2008}{\natexlab{b}}),
  \eprint{0807.3279}.

\bibitem[{\citenamefont{Essig et~al.}(2012{\natexlab{a}})\citenamefont{Essig,
  Mardon, and Volansky}}]{Essig:2011nj}
\bibinfo{author}{\bibfnamefont{R.}~\bibnamefont{Essig}},
  \bibinfo{author}{\bibfnamefont{J.}~\bibnamefont{Mardon}}, \bibnamefont{and}
  \bibinfo{author}{\bibfnamefont{T.}~\bibnamefont{Volansky}},
  \bibinfo{journal}{Phys. Rev.} \textbf{\bibinfo{volume}{D85}},
  \bibinfo{pages}{076007} (\bibinfo{year}{2012}{\natexlab{a}}),
  \eprint{1108.5383}.

\bibitem[{\citenamefont{Essig et~al.}(2012{\natexlab{b}})\citenamefont{Essig,
  Manalaysay, Mardon, Sorensen, and Volansky}}]{Essig:2012yx}
\bibinfo{author}{\bibfnamefont{R.}~\bibnamefont{Essig}},
  \bibinfo{author}{\bibfnamefont{A.}~\bibnamefont{Manalaysay}},
  \bibinfo{author}{\bibfnamefont{J.}~\bibnamefont{Mardon}},
  \bibinfo{author}{\bibfnamefont{P.}~\bibnamefont{Sorensen}}, \bibnamefont{and}
  \bibinfo{author}{\bibfnamefont{T.}~\bibnamefont{Volansky}},
  \bibinfo{journal}{Phys. Rev. Lett.} \textbf{\bibinfo{volume}{109}},
  \bibinfo{pages}{021301} (\bibinfo{year}{2012}{\natexlab{b}}),
  \eprint{1206.2644}.

\bibitem[{\citenamefont{An et~al.}(2015)\citenamefont{An, Pospelov, Pradler,
  and Ritz}}]{An:2014twa}
\bibinfo{author}{\bibfnamefont{H.}~\bibnamefont{An}},
  \bibinfo{author}{\bibfnamefont{M.}~\bibnamefont{Pospelov}},
  \bibinfo{author}{\bibfnamefont{J.}~\bibnamefont{Pradler}}, \bibnamefont{and}
  \bibinfo{author}{\bibfnamefont{A.}~\bibnamefont{Ritz}},
  \bibinfo{journal}{Phys. Lett.} \textbf{\bibinfo{volume}{B747}},
  \bibinfo{pages}{331} (\bibinfo{year}{2015}), \eprint{1412.8378}.

\bibitem[{\citenamefont{Lee et~al.}(2015)\citenamefont{Lee, Lisanti,
  Mishra-Sharma, and Safdi}}]{Lee:2015qva}
\bibinfo{author}{\bibfnamefont{S.~K.} \bibnamefont{Lee}},
  \bibinfo{author}{\bibfnamefont{M.}~\bibnamefont{Lisanti}},
  \bibinfo{author}{\bibfnamefont{S.}~\bibnamefont{Mishra-Sharma}},
  \bibnamefont{and} \bibinfo{author}{\bibfnamefont{B.~R.} \bibnamefont{Safdi}},
  \bibinfo{journal}{Phys. Rev.} \textbf{\bibinfo{volume}{D92}},
  \bibinfo{pages}{083517} (\bibinfo{year}{2015}), \eprint{1508.07361}.

\bibitem[{\citenamefont{Bloch et~al.}(2017)\citenamefont{Bloch, Essig, Tobioka,
  Volansky, and Yu}}]{Bloch:2016sjj}
\bibinfo{author}{\bibfnamefont{I.~M.} \bibnamefont{Bloch}},
  \bibinfo{author}{\bibfnamefont{R.}~\bibnamefont{Essig}},
  \bibinfo{author}{\bibfnamefont{K.}~\bibnamefont{Tobioka}},
  \bibinfo{author}{\bibfnamefont{T.}~\bibnamefont{Volansky}}, \bibnamefont{and}
  \bibinfo{author}{\bibfnamefont{T.-T.} \bibnamefont{Yu}},
  \bibinfo{journal}{JHEP} \textbf{\bibinfo{volume}{06}}, \bibinfo{pages}{087}
  (\bibinfo{year}{2017}), \eprint{1608.02123}.

\bibitem[{\citenamefont{Angloher et~al.}(2016)}]{Angloher:2015ewa}
\bibinfo{author}{\bibfnamefont{G.}~\bibnamefont{Angloher}} \bibnamefont{et~al.}
  (\bibinfo{collaboration}{CRESST}), \bibinfo{journal}{Eur. Phys. J.}
  \textbf{\bibinfo{volume}{C76}}, \bibinfo{pages}{25} (\bibinfo{year}{2016}),
  \eprint{1509.01515}.

\bibitem[{\citenamefont{Angloher et~al.}(2015)}]{Angloher:2015eza}
\bibinfo{author}{\bibfnamefont{G.}~\bibnamefont{Angloher}} \bibnamefont{et~al.}
  (\bibinfo{collaboration}{CRESST}) (\bibinfo{year}{2015}),
  \eprint{1503.08065}.

\bibitem[{\citenamefont{Aguilar-Arevalo
  et~al.}(2016)}]{Aguilar-Arevalo:2016ndq}
\bibinfo{author}{\bibfnamefont{A.}~\bibnamefont{Aguilar-Arevalo}}
  \bibnamefont{et~al.} (\bibinfo{collaboration}{DAMIC}),
  \bibinfo{journal}{Phys. Rev.} \textbf{\bibinfo{volume}{D94}},
  \bibinfo{pages}{082006} (\bibinfo{year}{2016}), \eprint{1607.07410}.

\bibitem[{\citenamefont{Hehn et~al.}(2016)}]{Hehn:2016nll}
\bibinfo{author}{\bibfnamefont{L.}~\bibnamefont{Hehn}} \bibnamefont{et~al.}
  (\bibinfo{collaboration}{EDELWEISS}), \bibinfo{journal}{Eur. Phys. J.}
  \textbf{\bibinfo{volume}{C76}}, \bibinfo{pages}{548} (\bibinfo{year}{2016}),
  \eprint{1607.03367}.

\bibitem[{\citenamefont{Agnese et~al.}(2016)}]{Agnese:2015nto}
\bibinfo{author}{\bibfnamefont{R.}~\bibnamefont{Agnese}} \bibnamefont{et~al.}
  (\bibinfo{collaboration}{SuperCDMS}), \bibinfo{journal}{Phys. Rev. Lett.}
  \textbf{\bibinfo{volume}{116}}, \bibinfo{pages}{071301}
  (\bibinfo{year}{2016}), \eprint{1509.02448}.

\bibitem[{\citenamefont{Agnese et~al.}(2017)}]{Agnese:2016cpb}
\bibinfo{author}{\bibfnamefont{R.}~\bibnamefont{Agnese}} \bibnamefont{et~al.}
  (\bibinfo{collaboration}{SuperCDMS}), \bibinfo{journal}{Phys. Rev.}
  \textbf{\bibinfo{volume}{D95}}, \bibinfo{pages}{082002}
  (\bibinfo{year}{2017}), \eprint{1610.00006}.

\bibitem[{\citenamefont{Kouvaris and Pradler}(2017)}]{Kouvaris:2016afs}
\bibinfo{author}{\bibfnamefont{C.}~\bibnamefont{Kouvaris}} \bibnamefont{and}
  \bibinfo{author}{\bibfnamefont{J.}~\bibnamefont{Pradler}},
  \bibinfo{journal}{Phys. Rev. Lett.} \textbf{\bibinfo{volume}{118}},
  \bibinfo{pages}{031803} (\bibinfo{year}{2017}), \eprint{1607.01789}.

\bibitem[{\citenamefont{Angle et~al.}(2011)}]{Angle:2011th}
\bibinfo{author}{\bibfnamefont{J.}~\bibnamefont{Angle}} \bibnamefont{et~al.}
  (\bibinfo{collaboration}{XENON10}), \bibinfo{journal}{Phys. Rev. Lett.}
  \textbf{\bibinfo{volume}{107}}, \bibinfo{pages}{051301}
  (\bibinfo{year}{2011}), \bibinfo{note}{[Erratum: Phys. Rev.
  Lett.110,249901(2013)]}, \eprint{1104.3088}.

\bibitem[{\citenamefont{Aprile et~al.}(2016{\natexlab{a}})}]{Aprile:2016wwo}
\bibinfo{author}{\bibfnamefont{E.}~\bibnamefont{Aprile}} \bibnamefont{et~al.}
  (\bibinfo{collaboration}{XENON}), \bibinfo{journal}{Phys. Rev.}
  \textbf{\bibinfo{volume}{D94}}, \bibinfo{pages}{092001}
  (\bibinfo{year}{2016}{\natexlab{a}}), \eprint{1605.06262}.

\bibitem[{\citenamefont{Akerib et~al.}(2016{\natexlab{a}})}]{Akerib:2015rjg}
\bibinfo{author}{\bibfnamefont{D.~S.} \bibnamefont{Akerib}}
  \bibnamefont{et~al.} (\bibinfo{collaboration}{LUX}), \bibinfo{journal}{Phys.
  Rev. Lett.} \textbf{\bibinfo{volume}{116}}, \bibinfo{pages}{161301}
  (\bibinfo{year}{2016}{\natexlab{a}}), \eprint{1512.03506}.

\bibitem[{\citenamefont{Akerib et~al.}(2017{\natexlab{a}})}]{Akerib:2016vxi}
\bibinfo{author}{\bibfnamefont{D.~S.} \bibnamefont{Akerib}}
  \bibnamefont{et~al.} (\bibinfo{collaboration}{LUX}), \bibinfo{journal}{Phys.
  Rev. Lett.} \textbf{\bibinfo{volume}{118}}, \bibinfo{pages}{021303}
  (\bibinfo{year}{2017}{\natexlab{a}}), \eprint{1608.07648}.

\bibitem[{\citenamefont{Marrodan~Undagoitia and
  Rauch}(2016)}]{Undagoitia:2015gya}
\bibinfo{author}{\bibfnamefont{T.}~\bibnamefont{Marrodan~Undagoitia}}
  \bibnamefont{and} \bibinfo{author}{\bibfnamefont{L.}~\bibnamefont{Rauch}},
  \bibinfo{journal}{J. Phys.} \textbf{\bibinfo{volume}{G43}},
  \bibinfo{pages}{013001} (\bibinfo{year}{2016}), \eprint{1509.08767}.

\bibitem[{\citenamefont{Akerib et~al.}(2015)}]{Akerib:2015cja}
\bibinfo{author}{\bibfnamefont{D.~S.} \bibnamefont{Akerib}}
  \bibnamefont{et~al.} (\bibinfo{collaboration}{LZ}) (\bibinfo{year}{2015}),
  \eprint{1509.02910}.

\bibitem[{\citenamefont{McCabe}(2014)}]{McCabe:2013kea}
\bibinfo{author}{\bibfnamefont{C.}~\bibnamefont{McCabe}},
  \bibinfo{journal}{JCAP} \textbf{\bibinfo{volume}{1402}}, \bibinfo{pages}{027}
  (\bibinfo{year}{2014}), \eprint{1312.1355}.

\bibitem[{\citenamefont{Lee et~al.}(2013)\citenamefont{Lee, Lisanti, and
  Safdi}}]{Lee:2013xxa}
\bibinfo{author}{\bibfnamefont{S.~K.} \bibnamefont{Lee}},
  \bibinfo{author}{\bibfnamefont{M.}~\bibnamefont{Lisanti}}, \bibnamefont{and}
  \bibinfo{author}{\bibfnamefont{B.~R.} \bibnamefont{Safdi}},
  \bibinfo{journal}{JCAP} \textbf{\bibinfo{volume}{1311}}, \bibinfo{pages}{033}
  (\bibinfo{year}{2013}), \eprint{1307.5323}.

\bibitem[{\citenamefont{Bozorgnia et~al.}(2016)\citenamefont{Bozorgnia, Calore,
  Schaller, Lovell, Bertone, Frenk, Crain, Navarro, Schaye, and
  Theuns}}]{Bozorgnia:2016ogo}
\bibinfo{author}{\bibfnamefont{N.}~\bibnamefont{Bozorgnia}},
  \bibinfo{author}{\bibfnamefont{F.}~\bibnamefont{Calore}},
  \bibinfo{author}{\bibfnamefont{M.}~\bibnamefont{Schaller}},
  \bibinfo{author}{\bibfnamefont{M.}~\bibnamefont{Lovell}},
  \bibinfo{author}{\bibfnamefont{G.}~\bibnamefont{Bertone}},
  \bibinfo{author}{\bibfnamefont{C.~S.} \bibnamefont{Frenk}},
  \bibinfo{author}{\bibfnamefont{R.~A.} \bibnamefont{Crain}},
  \bibinfo{author}{\bibfnamefont{J.~F.} \bibnamefont{Navarro}},
  \bibinfo{author}{\bibfnamefont{J.}~\bibnamefont{Schaye}}, \bibnamefont{and}
  \bibinfo{author}{\bibfnamefont{T.}~\bibnamefont{Theuns}},
  \bibinfo{journal}{JCAP} \textbf{\bibinfo{volume}{1605}}, \bibinfo{pages}{024}
  (\bibinfo{year}{2016}), \eprint{1601.04707}.

\bibitem[{\citenamefont{Kelso et~al.}(2016)\citenamefont{Kelso, Savage,
  Valluri, Freese, Stinson, and Bailin}}]{Kelso:2016qqj}
\bibinfo{author}{\bibfnamefont{C.}~\bibnamefont{Kelso}},
  \bibinfo{author}{\bibfnamefont{C.}~\bibnamefont{Savage}},
  \bibinfo{author}{\bibfnamefont{M.}~\bibnamefont{Valluri}},
  \bibinfo{author}{\bibfnamefont{K.}~\bibnamefont{Freese}},
  \bibinfo{author}{\bibfnamefont{G.~S.} \bibnamefont{Stinson}},
  \bibnamefont{and} \bibinfo{author}{\bibfnamefont{J.}~\bibnamefont{Bailin}},
  \bibinfo{journal}{JCAP} \textbf{\bibinfo{volume}{1608}}, \bibinfo{pages}{071}
  (\bibinfo{year}{2016}), \eprint{1601.04725}.

\bibitem[{\citenamefont{Sloane et~al.}(2016)\citenamefont{Sloane, Buckley,
  Brooks, and Governato}}]{Sloane:2016kyi}
\bibinfo{author}{\bibfnamefont{J.~D.} \bibnamefont{Sloane}},
  \bibinfo{author}{\bibfnamefont{M.~R.} \bibnamefont{Buckley}},
  \bibinfo{author}{\bibfnamefont{A.~M.} \bibnamefont{Brooks}},
  \bibnamefont{and}
  \bibinfo{author}{\bibfnamefont{F.}~\bibnamefont{Governato}},
  \bibinfo{journal}{ApJ} \textbf{\bibinfo{volume}{831}}, \bibinfo{eid}{93}
  (\bibinfo{year}{2016}), \eprint{1601.05402}.

\bibitem[{\citenamefont{Collar and Avignone~III}(1992)}]{Collar:1992qc}
\bibinfo{author}{\bibfnamefont{J.~I.} \bibnamefont{Collar}} \bibnamefont{and}
  \bibinfo{author}{\bibfnamefont{F.~T.} \bibnamefont{Avignone~III}},
  \bibinfo{journal}{Phys. Lett.} \textbf{\bibinfo{volume}{B275}},
  \bibinfo{pages}{181} (\bibinfo{year}{1992}).

\bibitem[{\citenamefont{Collar and Avignone~III}(1993)}]{Collar:1993ss}
\bibinfo{author}{\bibfnamefont{J.~I.} \bibnamefont{Collar}} \bibnamefont{and}
  \bibinfo{author}{\bibfnamefont{F.~T.} \bibnamefont{Avignone~III}},
  \bibinfo{journal}{Phys. Rev.} \textbf{\bibinfo{volume}{D47}},
  \bibinfo{pages}{5238} (\bibinfo{year}{1993}).

\bibitem[{\citenamefont{Kavanagh et~al.}(2017)\citenamefont{Kavanagh, Catena,
  and Kouvaris}}]{Kavanagh:2016pyr}
\bibinfo{author}{\bibfnamefont{B.~J.} \bibnamefont{Kavanagh}},
  \bibinfo{author}{\bibfnamefont{R.}~\bibnamefont{Catena}}, \bibnamefont{and}
  \bibinfo{author}{\bibfnamefont{C.}~\bibnamefont{Kouvaris}},
  \bibinfo{journal}{JCAP} \textbf{\bibinfo{volume}{1701}}, \bibinfo{pages}{012}
  (\bibinfo{year}{2017}), \eprint{1611.05453}.

\bibitem[{ato()}]{atomicff}
\bibinfo{note}{NIST database,
  \url{http://physics.nist.gov/PhysRefData/FFast/html/form.html} (Accessed 8
  Feb.\ 2017)}.

\bibitem[{\citenamefont{Vietze et~al.}(2015)\citenamefont{Vietze, Klos,
  Menendez, Haxton, and Schwenk}}]{Vietze:2014vsa}
\bibinfo{author}{\bibfnamefont{L.}~\bibnamefont{Vietze}},
  \bibinfo{author}{\bibfnamefont{P.}~\bibnamefont{Klos}},
  \bibinfo{author}{\bibfnamefont{J.}~\bibnamefont{Menendez}},
  \bibinfo{author}{\bibfnamefont{W.~C.} \bibnamefont{Haxton}},
  \bibnamefont{and} \bibinfo{author}{\bibfnamefont{A.}~\bibnamefont{Schwenk}},
  \bibinfo{journal}{Phys. Rev.} \textbf{\bibinfo{volume}{D91}},
  \bibinfo{pages}{043520} (\bibinfo{year}{2015}), \eprint{1412.6091}.

\bibitem[{\citenamefont{Chepel and Araujo}(2013)}]{Chepel:2012sj}
\bibinfo{author}{\bibfnamefont{V.}~\bibnamefont{Chepel}} \bibnamefont{and}
  \bibinfo{author}{\bibfnamefont{H.}~\bibnamefont{Araujo}},
  \bibinfo{journal}{JINST} \textbf{\bibinfo{volume}{8}},
  \bibinfo{pages}{R04001} (\bibinfo{year}{2013}), \eprint{1207.2292}.

\bibitem[{\citenamefont{Szydagis et~al.}(2011)\citenamefont{Szydagis, Barry,
  Kazkaz, Mock, Stolp, Sweany, Tripathi, Uvarov, Walsh, and
  Woods}}]{Szydagis:2011tk}
\bibinfo{author}{\bibfnamefont{M.}~\bibnamefont{Szydagis}},
  \bibinfo{author}{\bibfnamefont{N.}~\bibnamefont{Barry}},
  \bibinfo{author}{\bibfnamefont{K.}~\bibnamefont{Kazkaz}},
  \bibinfo{author}{\bibfnamefont{J.}~\bibnamefont{Mock}},
  \bibinfo{author}{\bibfnamefont{D.}~\bibnamefont{Stolp}},
  \bibinfo{author}{\bibfnamefont{M.}~\bibnamefont{Sweany}},
  \bibinfo{author}{\bibfnamefont{M.}~\bibnamefont{Tripathi}},
  \bibinfo{author}{\bibfnamefont{S.}~\bibnamefont{Uvarov}},
  \bibinfo{author}{\bibfnamefont{N.}~\bibnamefont{Walsh}}, \bibnamefont{and}
  \bibinfo{author}{\bibfnamefont{M.}~\bibnamefont{Woods}},
  \bibinfo{journal}{JINST} \textbf{\bibinfo{volume}{6}},
  \bibinfo{pages}{P10002} (\bibinfo{year}{2011}), \eprint{1106.1613}.

\bibitem[{\citenamefont{Szydagis et~al.}(2013)\citenamefont{Szydagis, Fyhrie,
  Thorngren, and Tripathi}}]{Szydagis:2013sih}
\bibinfo{author}{\bibfnamefont{M.}~\bibnamefont{Szydagis}},
  \bibinfo{author}{\bibfnamefont{A.}~\bibnamefont{Fyhrie}},
  \bibinfo{author}{\bibfnamefont{D.}~\bibnamefont{Thorngren}},
  \bibnamefont{and} \bibinfo{author}{\bibfnamefont{M.}~\bibnamefont{Tripathi}},
  \bibinfo{journal}{JINST} \textbf{\bibinfo{volume}{8}},
  \bibinfo{pages}{C10003} (\bibinfo{year}{2013}), \eprint{1307.6601}.

\bibitem[{\citenamefont{Lenardo et~al.}(2015)\citenamefont{Lenardo, Kazkaz,
  Manalaysay, Mock, Szydagis, and Tripathi}}]{Lenardo:2014cva}
\bibinfo{author}{\bibfnamefont{B.}~\bibnamefont{Lenardo}},
  \bibinfo{author}{\bibfnamefont{K.}~\bibnamefont{Kazkaz}},
  \bibinfo{author}{\bibfnamefont{A.}~\bibnamefont{Manalaysay}},
  \bibinfo{author}{\bibfnamefont{J.}~\bibnamefont{Mock}},
  \bibinfo{author}{\bibfnamefont{M.}~\bibnamefont{Szydagis}}, \bibnamefont{and}
  \bibinfo{author}{\bibfnamefont{M.}~\bibnamefont{Tripathi}},
  \bibinfo{journal}{IEEE Trans. Nucl. Sci.} \textbf{\bibinfo{volume}{62}},
  \bibinfo{pages}{3387} (\bibinfo{year}{2015}), \eprint{1412.4417}.

\bibitem[{\citenamefont{Sorensen}(2015)}]{Sorensen:2014sla}
\bibinfo{author}{\bibfnamefont{P.}~\bibnamefont{Sorensen}},
  \bibinfo{journal}{Phys. Rev.} \textbf{\bibinfo{volume}{D91}},
  \bibinfo{pages}{083509} (\bibinfo{year}{2015}), \eprint{1412.3028}.

\bibitem[{\citenamefont{Verbus et~al.}(2017)}]{Verbus:2016sgw}
\bibinfo{author}{\bibfnamefont{J.~R.} \bibnamefont{Verbus}}
  \bibnamefont{et~al.}, \bibinfo{journal}{Nucl. Instrum. Meth.}
  \textbf{\bibinfo{volume}{A851}}, \bibinfo{pages}{68} (\bibinfo{year}{2017}),
  \eprint{1608.05309}.

\bibitem[{\citenamefont{Akerib et~al.}(2016{\natexlab{b}})}]{Akerib:2015wdi}
\bibinfo{author}{\bibfnamefont{D.~S.} \bibnamefont{Akerib}}
  \bibnamefont{et~al.} (\bibinfo{collaboration}{LUX}), \bibinfo{journal}{Phys.
  Rev.} \textbf{\bibinfo{volume}{D93}}, \bibinfo{pages}{072009}
  (\bibinfo{year}{2016}{\natexlab{b}}), \eprint{1512.03133}.

\bibitem[{xe1()}]{xe127}
\bibinfo{note}{D.~Huang, UCLA DM2016 presentation,
  \url{https://conferences.pa.ucla.edu/dm16/talks/huang.pdf} (Accessed 8 Feb.\
  2017)}.

\bibitem[{Bou()}]{BoultonAPS}
\bibinfo{note}{E.~Boulton, APS Meeting 2017 presentation,
  \url{https://meetings.aps.org/Meeting/APR17/Session/B13.7} (Accessed 28 Feb.\
  2017)}.

\bibitem[{\citenamefont{Boulton et~al.}(2017)}]{Boulton:2017hub}
\bibinfo{author}{\bibfnamefont{E.~M.} \bibnamefont{Boulton}}
  \bibnamefont{et~al.} (\bibinfo{year}{2017}), \eprint{1705.08958}.

\bibitem[{\citenamefont{Akerib et~al.}(2017{\natexlab{b}})}]{Akerib:2016qlr}
\bibinfo{author}{\bibfnamefont{D.~S.} \bibnamefont{Akerib}}
  \bibnamefont{et~al.} (\bibinfo{collaboration}{LUX}), \bibinfo{journal}{Phys.
  Rev.} \textbf{\bibinfo{volume}{D95}}, \bibinfo{pages}{012008}
  (\bibinfo{year}{2017}{\natexlab{b}}), \eprint{1610.02076}.

\bibitem[{\citenamefont{Akerib et~al.}(2016{\natexlab{c}})}]{Akerib:2016mzi}
\bibinfo{author}{\bibfnamefont{D.~S.} \bibnamefont{Akerib}}
  \bibnamefont{et~al.} (\bibinfo{collaboration}{LUX})
  (\bibinfo{year}{2016}{\natexlab{c}}), \eprint{1608.05381}.

\bibitem[{\citenamefont{Bailey}(2016)}]{Bailey}
\bibinfo{author}{\bibfnamefont{A.}~\bibnamefont{Bailey}}, Ph.D. thesis,
  \bibinfo{school}{Imperial College London} (\bibinfo{year}{2016}).

\bibitem[{\citenamefont{Dobi}(2014)}]{Dobi}
\bibinfo{author}{\bibfnamefont{A.}~\bibnamefont{Dobi}}, Ph.D. thesis,
  \bibinfo{school}{University of Maryland} (\bibinfo{year}{2014}).

\bibitem[{\citenamefont{Akerib et~al.}(2016{\natexlab{d}})}]{Akerib:2016lao}
\bibinfo{author}{\bibfnamefont{D.~S.} \bibnamefont{Akerib}}
  \bibnamefont{et~al.} (\bibinfo{collaboration}{LUX}), \bibinfo{journal}{Phys.
  Rev. Lett.} \textbf{\bibinfo{volume}{116}}, \bibinfo{pages}{161302}
  (\bibinfo{year}{2016}{\natexlab{d}}), \eprint{1602.03489}.

\bibitem[{szy()}]{szydagisICHEP}
\bibinfo{note}{M.~Szydagis, ICHEP2016 presentation,
  \url{https://indico.cern.ch/event/432527/contributions/1071603/attachments/1321295/1981600/ICHEP2016_Szydagis.pdf}
  (Accessed 8 Feb.\ 2017)}.

\bibitem[{Sha()}]{ShawIDM}
\bibinfo{note}{S.~Shaw, IDM2016 presentation,
  \url{https://idm2016.shef.ac.uk/indico/event/0/session/5/contribution/29/material/slides/0.pdf}
  (Accessed 8 Feb.\ 2017)}.

\bibitem[{\citenamefont{Szydagis}(2016)}]{Szydagis:2016few}
\bibinfo{author}{\bibfnamefont{M.}~\bibnamefont{Szydagis}}
  (\bibinfo{collaboration}{LUX, LZ}), \bibinfo{journal}{PoS}
  \textbf{\bibinfo{volume}{ICHEP2016}}, \bibinfo{pages}{220}
  (\bibinfo{year}{2016}), \eprint{1611.05525}.

\bibitem[{\citenamefont{Akerib et~al.}(2017{\natexlab{c}})}]{Akerib:2017uem}
\bibinfo{author}{\bibfnamefont{D.~S.} \bibnamefont{Akerib}}
  \bibnamefont{et~al.} (\bibinfo{collaboration}{LUX}), \bibinfo{journal}{Phys.
  Rev. Lett.} \textbf{\bibinfo{volume}{118}}, \bibinfo{pages}{261301}
  (\bibinfo{year}{2017}{\natexlab{c}}), \eprint{1704.02297}.

\bibitem[{\citenamefont{Cowan et~al.}(2011)\citenamefont{Cowan, Cranmer, Gross,
  and Vitells}}]{Cowan:2010js}
\bibinfo{author}{\bibfnamefont{G.}~\bibnamefont{Cowan}},
  \bibinfo{author}{\bibfnamefont{K.}~\bibnamefont{Cranmer}},
  \bibinfo{author}{\bibfnamefont{E.}~\bibnamefont{Gross}}, \bibnamefont{and}
  \bibinfo{author}{\bibfnamefont{O.}~\bibnamefont{Vitells}},
  \bibinfo{journal}{Eur. Phys. J.} \textbf{\bibinfo{volume}{C71}},
  \bibinfo{pages}{1554} (\bibinfo{year}{2011}), \bibinfo{note}{[Erratum: Eur.
  Phys. J.C73,2501(2013)]}, \eprint{1007.1727}.

\bibitem[{\citenamefont{Barlow}(1990)}]{Barlow:1990vc}
\bibinfo{author}{\bibfnamefont{R.~J.} \bibnamefont{Barlow}},
  \bibinfo{journal}{Nucl. Instrum. Meth.} \textbf{\bibinfo{volume}{A297}},
  \bibinfo{pages}{496} (\bibinfo{year}{1990}).

\bibitem[{\citenamefont{Priel et~al.}(2017)\citenamefont{Priel, Rauch,
  Landsman, Manfredini, and Budnik}}]{Priel:2016apy}
\bibinfo{author}{\bibfnamefont{N.}~\bibnamefont{Priel}},
  \bibinfo{author}{\bibfnamefont{L.}~\bibnamefont{Rauch}},
  \bibinfo{author}{\bibfnamefont{H.}~\bibnamefont{Landsman}},
  \bibinfo{author}{\bibfnamefont{A.}~\bibnamefont{Manfredini}},
  \bibnamefont{and} \bibinfo{author}{\bibfnamefont{R.}~\bibnamefont{Budnik}},
  \bibinfo{journal}{JCAP} \textbf{\bibinfo{volume}{1705}}, \bibinfo{pages}{013}
  (\bibinfo{year}{2017}), \eprint{1610.02643}.

\bibitem[{\citenamefont{Mack et~al.}(2007)\citenamefont{Mack, Beacom, and
  Bertone}}]{Mack:2007xj}
\bibinfo{author}{\bibfnamefont{G.~D.} \bibnamefont{Mack}},
  \bibinfo{author}{\bibfnamefont{J.~F.} \bibnamefont{Beacom}},
  \bibnamefont{and} \bibinfo{author}{\bibfnamefont{G.}~\bibnamefont{Bertone}},
  \bibinfo{journal}{Phys. Rev.} \textbf{\bibinfo{volume}{D76}},
  \bibinfo{pages}{043523} (\bibinfo{year}{2007}), \eprint{0705.4298}.

\bibitem[{\citenamefont{Angloher et~al.}(2017)}]{Angloher:2017zkf}
\bibinfo{author}{\bibfnamefont{G.}~\bibnamefont{Angloher}} \bibnamefont{et~al.}
  (\bibinfo{collaboration}{CRESST}) (\bibinfo{year}{2017}),
  \eprint{1701.08157}.

\bibitem[{\citenamefont{Yellin}(2002)}]{Yellin:2002xd}
\bibinfo{author}{\bibfnamefont{S.}~\bibnamefont{Yellin}},
  \bibinfo{journal}{Phys. Rev.} \textbf{\bibinfo{volume}{D66}},
  \bibinfo{pages}{032005} (\bibinfo{year}{2002}), \eprint{physics/0203002}.

\bibitem[{\citenamefont{Aprile et~al.}(2016{\natexlab{b}})}]{Aprile:2015uzo}
\bibinfo{author}{\bibfnamefont{E.}~\bibnamefont{Aprile}} \bibnamefont{et~al.}
  (\bibinfo{collaboration}{XENON}), \bibinfo{journal}{JCAP}
  \textbf{\bibinfo{volume}{1604}}, \bibinfo{pages}{027}
  (\bibinfo{year}{2016}{\natexlab{b}}), \eprint{1512.07501}.

\bibitem[{\citenamefont{Cao et~al.}(2014)}]{Cao:2014jsa}
\bibinfo{author}{\bibfnamefont{X.}~\bibnamefont{Cao}} \bibnamefont{et~al.}
  (\bibinfo{collaboration}{PandaX}), \bibinfo{journal}{Sci. China Phys. Mech.
  Astron.} \textbf{\bibinfo{volume}{57}}, \bibinfo{pages}{1476}
  (\bibinfo{year}{2014}), \eprint{1405.2882}.

\bibitem[{\citenamefont{Mount et~al.}(2017)}]{Mount:2017qzi}
\bibinfo{author}{\bibfnamefont{B.~J.} \bibnamefont{Mount}} \bibnamefont{et~al.}
  (\bibinfo{year}{2017}), \eprint{1703.09144}.

\bibitem[{\citenamefont{Bergstrom et~al.}(2016)\citenamefont{Bergstrom,
  Gonzalez-Garcia, Maltoni, Pena-Garay, Serenelli, and
  Song}}]{Bergstrom:2016cbh}
\bibinfo{author}{\bibfnamefont{J.}~\bibnamefont{Bergstrom}},
  \bibinfo{author}{\bibfnamefont{M.~C.} \bibnamefont{Gonzalez-Garcia}},
  \bibinfo{author}{\bibfnamefont{M.}~\bibnamefont{Maltoni}},
  \bibinfo{author}{\bibfnamefont{C.}~\bibnamefont{Pena-Garay}},
  \bibinfo{author}{\bibfnamefont{A.~M.} \bibnamefont{Serenelli}},
  \bibnamefont{and} \bibinfo{author}{\bibfnamefont{N.}~\bibnamefont{Song}},
  \bibinfo{journal}{JHEP} \textbf{\bibinfo{volume}{03}}, \bibinfo{pages}{132}
  (\bibinfo{year}{2016}), \eprint{1601.00972}.

\bibitem[{\citenamefont{Baudis et~al.}(2014)\citenamefont{Baudis, Ferella,
  Kish, Manalaysay, Marrodan~Undagoitia, and Schumann}}]{Baudis:2013qla}
\bibinfo{author}{\bibfnamefont{L.}~\bibnamefont{Baudis}},
  \bibinfo{author}{\bibfnamefont{A.}~\bibnamefont{Ferella}},
  \bibinfo{author}{\bibfnamefont{A.}~\bibnamefont{Kish}},
  \bibinfo{author}{\bibfnamefont{A.}~\bibnamefont{Manalaysay}},
  \bibinfo{author}{\bibfnamefont{T.}~\bibnamefont{Marrodan~Undagoitia}},
  \bibnamefont{and} \bibinfo{author}{\bibfnamefont{M.}~\bibnamefont{Schumann}},
  \bibinfo{journal}{JCAP} \textbf{\bibinfo{volume}{1401}}, \bibinfo{pages}{044}
  (\bibinfo{year}{2014}), \eprint{1309.7024}.

\bibitem[{\citenamefont{Cerdeno et~al.}(2016)\citenamefont{Cerdeno, Fairbairn,
  Jubb, Machado, Vincent, and Boehm}}]{Cerdeno:2016sfi}
\bibinfo{author}{\bibfnamefont{D.~G.} \bibnamefont{Cerdeno}},
  \bibinfo{author}{\bibfnamefont{M.}~\bibnamefont{Fairbairn}},
  \bibinfo{author}{\bibfnamefont{T.}~\bibnamefont{Jubb}},
  \bibinfo{author}{\bibfnamefont{P.~A.~N.} \bibnamefont{Machado}},
  \bibinfo{author}{\bibfnamefont{A.~C.} \bibnamefont{Vincent}},
  \bibnamefont{and} \bibinfo{author}{\bibfnamefont{C.}~\bibnamefont{Boehm}},
  \bibinfo{journal}{JHEP} \textbf{\bibinfo{volume}{05}}, \bibinfo{pages}{118}
  (\bibinfo{year}{2016}), \bibinfo{note}{[Erratum: JHEP09,048(2016)]},
  \eprint{1604.01025}.

\bibitem[{\citenamefont{Chakraborty et~al.}(2014)\citenamefont{Chakraborty,
  Bhattacharjee, and Kar}}]{Chakraborty:2013zua}
\bibinfo{author}{\bibfnamefont{S.}~\bibnamefont{Chakraborty}},
  \bibinfo{author}{\bibfnamefont{P.}~\bibnamefont{Bhattacharjee}},
  \bibnamefont{and} \bibinfo{author}{\bibfnamefont{K.}~\bibnamefont{Kar}},
  \bibinfo{journal}{Phys. Rev.} \textbf{\bibinfo{volume}{D89}},
  \bibinfo{pages}{013011} (\bibinfo{year}{2014}), \eprint{1309.4492}.

\bibitem[{\citenamefont{Davis}(2016)}]{Davis:2016dqh}
\bibinfo{author}{\bibfnamefont{J.~H.} \bibnamefont{Davis}}
  (\bibinfo{year}{2016}), \eprint{1605.00011}.

\bibitem[{\citenamefont{Lang et~al.}(2016)\citenamefont{Lang, McCabe, Reichard,
  Selvi, and Tamborra}}]{Lang:2016zhv}
\bibinfo{author}{\bibfnamefont{R.~F.} \bibnamefont{Lang}},
  \bibinfo{author}{\bibfnamefont{C.}~\bibnamefont{McCabe}},
  \bibinfo{author}{\bibfnamefont{S.}~\bibnamefont{Reichard}},
  \bibinfo{author}{\bibfnamefont{M.}~\bibnamefont{Selvi}}, \bibnamefont{and}
  \bibinfo{author}{\bibfnamefont{I.}~\bibnamefont{Tamborra}},
  \bibinfo{journal}{Phys. Rev.} \textbf{\bibinfo{volume}{D94}},
  \bibinfo{pages}{103009} (\bibinfo{year}{2016}), \eprint{1606.09243}.

\bibitem[{\citenamefont{Hardy et~al.}(2015)\citenamefont{Hardy, Lasenby,
  March-Russell, and West}}]{Hardy:2015boa}
\bibinfo{author}{\bibfnamefont{E.}~\bibnamefont{Hardy}},
  \bibinfo{author}{\bibfnamefont{R.}~\bibnamefont{Lasenby}},
  \bibinfo{author}{\bibfnamefont{J.}~\bibnamefont{March-Russell}},
  \bibnamefont{and} \bibinfo{author}{\bibfnamefont{S.~M.} \bibnamefont{West}},
  \bibinfo{journal}{JHEP} \textbf{\bibinfo{volume}{07}}, \bibinfo{pages}{133}
  (\bibinfo{year}{2015}), \eprint{1504.05419}.

\bibitem[{\citenamefont{Butcher et~al.}(2016)\citenamefont{Butcher, Kirk,
  Monroe, and West}}]{Butcher:2016hic}
\bibinfo{author}{\bibfnamefont{A.}~\bibnamefont{Butcher}},
  \bibinfo{author}{\bibfnamefont{R.}~\bibnamefont{Kirk}},
  \bibinfo{author}{\bibfnamefont{J.}~\bibnamefont{Monroe}}, \bibnamefont{and}
  \bibinfo{author}{\bibfnamefont{S.~M.} \bibnamefont{West}}
  (\bibinfo{year}{2016}), \eprint{1610.01840}.

\bibitem[{\citenamefont{Cherry et~al.}(2015)\citenamefont{Cherry, Frandsen, and
  Shoemaker}}]{Cherry:2015oca}
\bibinfo{author}{\bibfnamefont{J.~F.} \bibnamefont{Cherry}},
  \bibinfo{author}{\bibfnamefont{M.~T.} \bibnamefont{Frandsen}},
  \bibnamefont{and} \bibinfo{author}{\bibfnamefont{I.~M.}
  \bibnamefont{Shoemaker}}, \bibinfo{journal}{Phys. Rev. Lett.}
  \textbf{\bibinfo{volume}{114}}, \bibinfo{pages}{231303}
  (\bibinfo{year}{2015}), \eprint{1501.03166}.

\bibitem[{\citenamefont{Baudis et~al.}(2013)\citenamefont{Baudis, Kessler,
  Klos, Lang, Menendez, Reichard, and Schwenk}}]{Baudis:2013bba}
\bibinfo{author}{\bibfnamefont{L.}~\bibnamefont{Baudis}},
  \bibinfo{author}{\bibfnamefont{G.}~\bibnamefont{Kessler}},
  \bibinfo{author}{\bibfnamefont{P.}~\bibnamefont{Klos}},
  \bibinfo{author}{\bibfnamefont{R.~F.} \bibnamefont{Lang}},
  \bibinfo{author}{\bibfnamefont{J.}~\bibnamefont{Menendez}},
  \bibinfo{author}{\bibfnamefont{S.}~\bibnamefont{Reichard}}, \bibnamefont{and}
  \bibinfo{author}{\bibfnamefont{A.}~\bibnamefont{Schwenk}},
  \bibinfo{journal}{Phys. Rev.} \textbf{\bibinfo{volume}{D88}},
  \bibinfo{pages}{115014} (\bibinfo{year}{2013}), \eprint{1309.0825}.

\bibitem[{\citenamefont{McCabe}(2016)}]{McCabe:2015eia}
\bibinfo{author}{\bibfnamefont{C.}~\bibnamefont{McCabe}},
  \bibinfo{journal}{JCAP} \textbf{\bibinfo{volume}{1605}}, \bibinfo{pages}{033}
  (\bibinfo{year}{2016}), \eprint{1512.00460}.

\bibitem[{\citenamefont{Thomas and Imel}(1987)}]{Thomas:1987zz}
\bibinfo{author}{\bibfnamefont{J.}~\bibnamefont{Thomas}} \bibnamefont{and}
  \bibinfo{author}{\bibfnamefont{D.~A.} \bibnamefont{Imel}},
  \bibinfo{journal}{Phys. Rev.} \textbf{\bibinfo{volume}{A36}},
  \bibinfo{pages}{614} (\bibinfo{year}{1987}).

\bibitem[{\citenamefont{Goetzke et~al.}(2016)\citenamefont{Goetzke, Aprile,
  Anthony, Plante, and Weber}}]{Goetzke:2016lfg}
\bibinfo{author}{\bibfnamefont{L.~W.} \bibnamefont{Goetzke}},
  \bibinfo{author}{\bibfnamefont{E.}~\bibnamefont{Aprile}},
  \bibinfo{author}{\bibfnamefont{M.}~\bibnamefont{Anthony}},
  \bibinfo{author}{\bibfnamefont{G.}~\bibnamefont{Plante}}, \bibnamefont{and}
  \bibinfo{author}{\bibfnamefont{M.}~\bibnamefont{Weber}}
  (\bibinfo{year}{2016}), \eprint{1611.10322}.

\bibitem[{\citenamefont{Schumann et~al.}(2015)\citenamefont{Schumann, Baudis,
  Butikofer, Kish, and Selvi}}]{Schumann:2015cpa}
\bibinfo{author}{\bibfnamefont{M.}~\bibnamefont{Schumann}},
  \bibinfo{author}{\bibfnamefont{L.}~\bibnamefont{Baudis}},
  \bibinfo{author}{\bibfnamefont{L.}~\bibnamefont{Butikofer}},
  \bibinfo{author}{\bibfnamefont{A.}~\bibnamefont{Kish}}, \bibnamefont{and}
  \bibinfo{author}{\bibfnamefont{M.}~\bibnamefont{Selvi}},
  \bibinfo{journal}{JCAP} \textbf{\bibinfo{volume}{1510}}, \bibinfo{pages}{016}
  (\bibinfo{year}{2015}), \eprint{1506.08309}.

\end{thebibliography}
\end{document}